\documentclass[runningheads]{llncs}

\usepackage{graphicx,xcolor,url,booktabs,amssymb,amsmath}

\usepackage{mystyle_llncs-compatible}

\titlerunning{Distance Oracles for Time-Dependent Networks}
\authorrunning{S.~Kontogiannis and C.~Zaroliagis}

\title{Distance Oracles for Time-Dependent Networks\thanks{%
		\vskip-9pt\noindent
		This work was supported by EU FP7/2007-2013 under grant agreements no.~288094 (eCOMPASS) and no.~609026 (MOVESMART),
		and partially done while both authors were visiting
		the Karlsruhe~Instute~of~Technology.
		}
}

\author{%
	Spyros Kontogiannis\inst{1,2}
	\and
	Christos Zaroliagis\inst{2,3}
}

\institute{%
    Dept.~of Comp.~Science \& Engineering, U. Ioannina, 45110 Ioannina, Greece \email{kontog@cs.uoi.gr}%
	\vskip3pt
	\and
    Computer Technology Institute \& Press ``Diophantus'',
	26504 Patras, Greece
	\vskip3pt
	\and
	Dept.~of Comp.~Engineering \& Informatics, U. Patras, 26504 Patras, Greece
	\email{zaro@ceid.upatras.gr}%
    	\\[5pt]
      		\today
	}

\begin{document}

\maketitle

\begin{abstract}

We present the first approximate distance oracle for sparse directed networks with \emph{time-dependent} arc-travel-times determined by continuous, piecewise linear, positive functions possessing the FIFO property.
Our approach precomputes $(1+\eps)-$approximate distance summaries from selected landmark vertices to all other vertices in the network.
Our oracle uses subquadratic space and time preprocessing, and provides two sublinear-time query algorithms that deliver constant and $(1+\sigma)-$approximate shortest-travel-times, respectively, for arbitrary origin-destination pairs in the network, for any constant $\sigma > \eps$. Our oracle is based only on the sparsity of the network, along with two quite natural assumptions about travel-time functions which allow the smooth transition towards asymmetric and time-dependent distance metrics.

\end{abstract}


\section{Introduction}
\label{section:intro}

Distance oracles are succinct data structures encoding shortest path information among a carefully selected subset of pairs of vertices in a graph. The encoding is done in such a way that the oracle can efficiently answer shortest path queries for arbitrary origin-destination pairs, exploiting the preprocessed data and/or local shortest path searches. A distance oracle is exact (resp.~approximate) if the returned distances by the accompanying query algorithm are exact (resp.~approximate).
A bulk of important work (e.g.,
\cite{Agarwal-Godfrey-2013,Patrascu-Roditty-R2010,Porat-Roditty-2011,Sommer-Verbin-Yu-2009,Thorup-Zwick-2005,Wulf-Nilsen-2012a,Wulf-Nilsen-2012b})
is devoted to constructing distance oracles for \emph{static} (i.e., \emph{time-independent}), mostly undirected networks in which the arc-costs are fixed, providing trade-offs between the oracle's space and query time and, in case of approximate oracles, also of the stretch (maximum ratio, over all origin-destination pairs, between the distance returned by the oracle and the actual distance). For an overview of distance oracles for static networks, the reader is referred to \cite{2014-Sommer-spq-survey} and references therein.

\subsection{Problem setting and motivation}

In many real-world applications, the arc costs may vary as functions of time (e.g., when representing travel-times) giving rise to \emph{time-dependent} network models. A striking example is route planning in road networks where the travel-time for traversing an arc $a=uv$ (modelling a road segment) depends on the temporal traffic conditions while traversing $uv$, and thus on the departure time from its tail $u$. Consequently, the optimal origin-destination path may vary with the departure-time from the origin. Apart from the theoretical challenge, the time-dependent model is also much more appropriate with respect to the historic traffic data that the route planning vendors have to digest, in order to provide their customers with fast route plans.
To see why it is more appropriate, consider, for example, TomTom's \emph{LiveTraffic} service\footnote{\url{http://www.tomtom.com/livetraffic/}}
which provides real-time estimations of average travel-time values, collected by periodically sampling the average speed of each road segment in a city,
using the cars connected to the service as sampling devices.
The crux is how to exploit all this historic traffic information, in order to \emph{efficiently} provide route plans that will adapt to the departure-time from the origin.
A customary way towards this direction is to consider the continuous piecewise linear (pwl) interpolants of
these sample points as \emph{arc-travel-time functions} of the corresponding instance.

Computing a time-dependent shortest path for a triple $(o,d,t_o)$ of an origin $o$, a destination $d$ and a departure-time $t_o$ from the origin,
has been studied extensively (see e.g., \cite{Cooke-Halsey-1966,Dreyfus1969,or-spmda-90}). The shape of arc-travel-time functions and the waiting policy at vertices may considerably affect the tractability of the problem \cite{or-spmda-90}. A crucial property is the \emph{FIFO property}, according to which each arc-arrival-time at the head of an arc is a \emph{non-decreasing} function of the departure-time from the tail.
If \emph{waiting-at-vertices} is forbidden and the arc-travel-time functions may be non-FIFO, then subpath optimality and simplicity of shortest paths is not guaranteed  \cite{or-spmda-90}. Thus, even if it exists, an optimal route is not computable by (extensions of) well known techniques, such as Dijkstra or Bellman-Ford. Additionally, many variants of the problem are also $\ComplClass{NP}-$hard \cite{SOS1998}.
On the other hand, if arc-travel-time functions possess the FIFO property, then the problem can be solved in polynomial time by a straightforward variant of Dijkstra's algorithm ($\alg{TDD}$), which relaxes arcs by computing the arc costs ``on the fly'', when scanning their tails. This has been first observed in \cite{Dreyfus1969}, where the \emph{unrestricted waiting policy} was (implicitly) assumed for vertices, along with the non-FIFO property for arcs.
%
The FIFO property may seem unreasonable in some application scenarios, e.g., when travellers at the dock of a train station wonder whether to take the very next slow train towards destination, or wait for a subsequent but faster train.

Our motivation in this work stems from \emph{route planning} in urban-traffic road networks where the FIFO property seems much more natural,
since all cars are assumed to travel according to the same (possibly time-dependent) average speed in each road segment,
and overtaking is not considered as an option when choosing a route plan.
Indeed, the raw traffic data for arc-travel-time functions by TomTom for the city of Berlin are compliant with this assumption \cite{ecompass}. Additionally, when shortest-travel-times are well defined and optimal waiting-times at nodes always exist, a non-FIFO arc with \emph{unrestricted-waiting-at-tail} policy is equivalent to a FIFO arc in which waiting at the tail is not beneficial \cite{or-spmda-90}.
Therefore, our focus in this work is on networks with FIFO arc-travel-time functions.

\subsection{Related work and main challenge}

The study of shortest paths is one of the cornerstone problems in Computer Science and Operations Research. Apart from the well-studied case of instances with static arc weights, several variants towards time-evolving instances have appeared in the literature.
We start by mentioning briefly the most characteristic attempts regarding non time-dependent models,
and subsequently we focus exclusively on related work regarding time-dependent shortest path models.

In the \emph{dynamic shortest path} problem (e.g., \cite{2005-Demetrescu-Italiano,1999-King,2004-Roditty-Zwick,2005-Thorup}), the arcs are allowed to be inserted to and/or deleted from the graph in an online fashion. The focus is on maintaining and efficiently updating a data structure representing the shortest path tree from a single source, or at least supporting fast shortest path queries between arbitrary vertices, in response to these changes. The main difference with the problem we study is exactly the online fashion of the changes in the characteristics of the graph metric.
%
In \emph{temporal networks} (e.g., \cite{2014-Akrida-Gasieniec-Mertzios-Spirakis,2002-Kempe-Kleinberg-Kumar,2013-Mertzios-Michail-Chatzigiannakis-Spirakis}), each arc comes with a vector of discrete arc-labels determining the time-slots of its availability. The goal is then to study the reachability and/or computation of shortest paths for arbitrary pairs of vertices, given that the chosen connecting path must also possess at least one non-decreasing subsequence of arc-labels, as we move from the origin to the destination. This problem is indeed a special case of the time-dependent shortest paths problem, in the sense that the availability patterns may be encoded as distance functions which switch between a finite and an infinite traversal cost. Typically these problems do not possess the FIFO property, but one may exploit the discretization of the time axis, which essentially determines the complexity of the instance to solve.
%
In the \emph{stochastic shortest path} problem (e.g., see \cite{1991-Bertsekas-Tsitsiklis,2006-Nikolova-Brand-Karger,2006-Nikolova-Kelner-Brand-Mitzenmacher}) the uncertainty of the arc weights is modeled by considering them as random variables. The goal is again the computation of paths with minimum expected weight. This is also a hard problem, but in the time-dependent shortest path there is no uncertainty on the behavior of the arcs.
%
In the \emph{parametric shortest path} problem, the graph comes with two distinct arc-weight vectors. The goal is to determine a shortest path with respect to any possible linear combination of the two weight vectors. It is well known \cite{2001-Mulmuley-Shah} that a shortest $od-$path may change $|V|^{\OmegaOrder{|V|}}$ times as the parameter of the linear combination changes. An upper bound of at most $|V|^{\Order{|V|}}$ is also well-known \cite{1980-Gusfield}. The main difference with the \emph{time-dependent shortest path} problem studied in the present work is that, when computing path lengths, rather than (essentially) composing the arrival-time functions of the constituent arcs, in the parametric shortest path problem
the arc-lengths are simply added.

Until recently, most of the previous work on the time-dependent shortest path problem concentrated on computing an optimal origin-destination path providing the earliest-arrival time at destination when departing at a \emph{given} time from the origin, and neglected the computational complexity of providing succinct representations of the entire earliest-arrival-time \emph{functions}, for \emph{all} departure-times from the origin.
Such representations, apart from allowing rapid answers to several queries for selected origin-destination pairs but for varying departure times, would also be valuable for the construction of \emph{distance summaries} (a.k.a. \emph{route planning maps}, or \emph{search profiles}) from central vertices (e.g., \emph{landmarks} or \emph{hubs}) towards other vertices in the network, providing a crucial ingredient for the construction of distance oracles to support real-time responses to arbitrary queries $(o,d,t_o)\in V\times V\times \reals$.

The complexity of succinctly representing earliest-arrival-time functions was first questioned in \cite{1999-Dean-MscThesis,2004-Dean-b,2004-Dean-a}, but was solved only recently by a seminal work \cite{Foschini-Hershberger-Suri-2011} which, for FIFO-abiding pwl arc-travel-time functions, showed that the problem of succinctly representing such a function for a \emph{single origin-destination pair} has space-complexity $(1+K)\cdot n^{\Theta(\log n)}$, where $n$ is the number of vertices and $K$ is the total number of breakpoints (or legs) of all the arc-travel-time functions.
Polynomial-time algorithms (or even PTAS) for constructing \emph{point-to-point} $(1+\varepsilon)$-approximate distance functions are provided in \cite{2010-Dehne-Omran-Sack,Foschini-Hershberger-Suri-2011},
delivering point-to-point travel-time values at most $1+\varepsilon$ times the true values.
Such approximate distance functions possess \emph{succinct representations}, since they require only $\Order{1+K}$ breakpoints per origin-destination pair. It is also easy to verify that $K$ could be substituted by the number $K^*$ of \emph{concavity-spoiling} breakpoints of the arc-travel-time functions (i.e., breakpoints at which the arc-travel-time slopes increase).

To the best of our knowledge, the problem of providing distance oracles for time-dependent networks with \emph{provably} good approximation guarantees, small preprocessing-space complexity and sublinear query time complexity, has not been investigated so far. Due to the hardness of providing succinct representations of exact shortest-travel-time functions, the only realistic alternative is to use approximations of these functions for the distance summaries that will be preprocessed and stored by the oracle. Exploiting a PTAS (such as that in \cite{Foschini-Hershberger-Suri-2011}) for computing approximate distance functions, one could provide a trivial oracle with query-time complexity $Q\in\Order{\log\log(K^*)}$, at the cost of an exceedingly high space-complexity $S\in\Order{(1+K^*)\cdot n^2}$, by storing succinct representations of all the point-to-point $(1+\eps)-$approximate shortest-travel-time functions.
At the other extreme, one might use the minimum possible space complexity $S\in \Order{n+m+K}$ for storing the input, at the cost of suffering a query-time complexity $Q\in \Order{m+n\log(n) [1+\log\log(1+K_{\max})]}$ (i.e., respond to each query by running $\alg{TDD}$ in real-time using a predecessor search structure for evaluating pwl functions)\footnote{$K_{\max}$ denotes the maximum number of breakpoints in an arc-travel-time function.}.
The main challenge considered in this work is to smoothly close the gap between these two extremes, i.e., to achieve a better (e.g., \emph{sublinear}) query-time complexity, while consuming smaller space-complexity (e.g., $\order{(1+K^*)\cdot n^2}$) for succinctly representing travel-time \emph{functions}, and enjoying a small (e.g., close to $1$) approximation guarantee (stretch factor).

\subsection{Our contribution}

We have successfully addressed the aforementioned challenge by
presenting the \emph{first} approximate distance oracle for sparse directed graphs with time-dependent arc-travel-times, which achieves all these goals. Our oracle is based only on the sparsity of the network, plus two assumptions of travel-time functions which are quite natural for route planning in road networks (cf.~Assumptions~\ref{assumption:Bounded-Travel-Time-Slopes} and \ref{assumption:Bounded-Opposite-Trips} in Section~\ref{section:preliminaries}).
It should be mentioned that:
(i) even in static undirected networks, achieving a stretch factor below $2$  using subquadratic space and sublinear query time, is possible only when $m\in \order{n^2}$, as it has been recently shown \cite{Agarwal-Godfrey-2013,Porat-Roditty-2011};
(ii) there is important applied work
\cite{2013-Batz-Geisberger-Sanders-Vetter,d-tdsr-11,dw-tdrp-09,ndls-bastd-12}
to develop time-dependent shortest path \emph{heuristics}, which however provide mainly empirical evidence on the success of the adopted approaches.

At a high level, our approach resembles the typical ones used in \emph{static} and \emph{undirected} graphs (e.g., \cite{Agarwal-Godfrey-2013,Porat-Roditty-2011,Thorup-Zwick-2005}): Distance summaries from selected landmarks
are precomputed and stored so as to support fast responses to arbitrary real-time queries
by growing small distance balls around the origin and the destination, and then closing
the gap between the prefix subpath from the origin and the suffix subpath towards the destination. However, it is not at all straightforward how this generic approach can be extended to \emph{time-dependent} and \emph{directed} graphs, since one is confronted with two highly non-trivial challenges: (i) handling directedness, and (ii) dealing with time-dependence, i.e., deciding the arrival-times to grow balls around vertices in the vicinity of the destination, because we simply do \textbf{\emph{not}} know the earliest-arrival-time at destination -- actually, this is what the original query to the oracle asks for.
A novelty of our query algorithms, contrary to other approaches, is exactly that we achieve the approximation guarantees by growing balls only from vertices around the origin. Managing this was a necessity for our analysis since growing balls around vertices in the vicinity of the destination at the \emph{right} arrival-time is essentially not an option.

Our specific contribution is as follows.
Let $U$ be the worst-case number of breakpoints for an $(1+\eps)-$approximation of a \emph{concave} distance function stored in our oracle,
and let $TDP$ be the maximum number of time-dependent shortest path probes required for their construction.
Then, we are able to construct a distance oracle that efficiently preprocess $(1+\eps)-$approximate distance functions from a set of landmarks, which are uniformly and independently selected with probability $\rho$, to all other vertices, in order to provide real-time responses to arbitrary queries via a recursive query algorithm of recursion depth (budget) $r$. The specific expected preprocessing
and query bounds of our oracle are presented in Table~\ref{table:tradeoffs} (3rd row) along with a comparison
with the best previous approaches (straightforward oracles).
\begin{table}[tbh]
\[
\begin{array}{|ll||c|c|c|}
\hline
\multicolumn{2}{|l||}{ \mbox{What is preprocessed} }
& \mbox{Preproc.~Space} 
& \mbox{Preproc.~Time} 
& \mbox{Query Time} 
\\
\hline\hline
\multicolumn{2}{|l||}{\mbox{All-To-All}}
& \Order{(K^*+1) n^2 U}
&  \Order{\begin{array}{c}n^2\log(n)  \\ \cdot \log\log(K_{\max}) \\   \cdot (K^*+1) TDP\end{array}}
& \Order{ \log\log(K^*) }
\\
\hline 
\multicolumn{2}{|l||}{\mbox{Nothing}}
& \Order{n+m+K}
&  \Order{1}
& \Order{	\begin{array}{c}
				n\log(n)\cdot
				\\
				\log\log(K_{\max})
			\end{array}
}
\\ \hline\hline
\multicolumn{2}{|l||}
{\begin{array}{c} \mbox{Landmarks-To-All} \\ \mbox{[This paper]} \end{array}}
& \Order{\rho n^2 (K^*+1) U}
&  \Order{\begin{array}{c}\rho n^2\log(n)  \\ \cdot \log\log(K_{\max}) \\   \cdot (K^*+1) TDP\end{array}}
& \Order{\begin{array}{c} \left(\frac{1}{\rho}\right)^{r+1} \cdot \log\left(\frac{1}{\rho}\right) \\  \cdot \log\log(K_{\max}) \end{array} }
\\ \hline
\multicolumn{2}{|l||}{%
\begin{array}{l}
K_{\max}\in\Order{1}
\\
\rho=n^{-a},
\\
U,TDP\in\Order{1}
\\
K^* \in \Order{\polylog(n)}
\end{array}}
& \tildeOrder{ n^{2-a} }
& \tildeOrder{ n^{2-a} }
& \tildeOrder{ n^{(r+1)a} }
\\ \hline
\end{array}
\]
\caption{\label{table:tradeoffs}%
Our main result (third row) and its comparison to the straightforward oracles with all-to-all preprocessing
and no preprocessing at all, for a given approximation guarantee $1+\eps$ of the preprocessed data.
The fourth row presents an explicit trade-off among preprocessing time/space and query time.
$\tildeOrder{~}$ hides polylogarithmic factors.
}
\label{table:tradeoffs}
\end{table}
Our oracle guarantees a stretch factor of $1 + \eps \frac{(1+\frac{\eps}{\psi})^{r+1}}{(1+\frac{\eps}{\psi})^{r+1}-1}$, where $\psi$ is a fixed constant depending on the characteristics of the arc-travel-time functions, but is independent of the network size. As it is proved in Theorem~\ref{thm:SO-BISECTION-performance} (Section~\ref{section:preprocessing}), $U$ and $TDP$ are independent of the network size $n$ and thus we can treat them as constants. Similarly, $K_{\max}$ (which is also part of the input) is considered to be independent
of the network size. But even if it was the case that $K_{\max}\in\ThetaOrder{K}$, this
would only have a doubly-logarithmic multiplicative effect in the preprocessing-time
and query-time complexities, which is indeed acceptable. Regarding the number $K^*$ of \emph{concavity-spoiling} breakpoints of arc-travel-time functions, note that if all arc-travel-time functions are concave, i.e., $K^*=0$, then we clearly achieve subquadratic preprocessing space and time for any $\rho\in\Order{n^{-\alpha}}$,
where $0 < \alpha < \frac{1}{r+1}$.
Real data (e.g., TomTom's traffic data for the city of Berlin~\cite{ecompass}) demonstrate that:
(i) only a small fraction of the arc-travel-time functions exhibit non-constant behavior;
(ii) for the vast majority of these non-constant-delay arcs, the arc-travel-time functions are either concave, or can be very tightly approximated by a typical \emph{concave} bell-shaped pwl function. It is thus only a tiny subset of critical arcs (e.g., bottleneck-segments in a large city) for which it would be indeed meaningful to consider also non-concave behavior. 
Our analysis guarantees that, when $K^*\in \order{n}$, one can fine-tune
the parameters of the oracles so that both sublinear query times and
subquadratic preprocessing space can be guaranteed. For example,
assuming $K^*\in\Order{\polylog(n)}$,
we get the trade-off presented in the 4th row of Table~\ref{table:tradeoffs}.
We also note that, apart from the choice of landmarks, our algorithms are deterministic.

The rest of the paper is organized as follows. Section~\ref{section:preliminaries} gives the ingredients and presents an overview of our approach. Section~\ref{section:preprocessing} presents our preprocessing algorithm. Our
constant approximation query algorithm is presented in Section~\ref{section:constant-approx-alg}, while our PTAS query algorithm is presented in  Section~\ref{section:ptas-alg}. Our main results are summarized in Section~\ref{section:main-results}.
The details on how to compute the actual path from the approximate distance values are presented in Section~\ref{sec:path-reconstruction}.
We conclude in Section~\ref{section:conlusions}. A preliminary version of this work appeared as \cite{KZ2014}.

\section{Ingredients and Overview of Our Approach}
\label{section:preliminaries}

\subsection{Notation}
\label{section:Notation}

Our input is provided by a network (directed graph) $G=(V,A)$ with $n$ vertices and $m=O(n)$ arcs.
Every arc $uv\in A$ is equipped with a periodic, continuous, piecewise-linear (pwl) \emph{arc-travel-time} (a.k.a. \emph{arc-delay}) function $D[uv]:\reals\rightarrow\positivereals$, such that
\(
	\forall k\in\integers, \forall t_u\in[0,T),~
		D[uv](k\cdot T + t_u) = D[uv](t_u)
\)
is the arc-travel-time of $uv$ when the departure-time from $u$ is $k\cdot T + t_u$. $D[uv]$ is represented succinctly as a continuous pwl function, by $K_{uv}$ breakpoints describing its projection to $[0,T)$. $K=\sum_{uv\in A} K_{uv}$ is the number of breakpoints to represent all the arc-delay functions in the network, and $K_{\max} = \max_{uv\in A} K_{uv}$. $K^*$ is the number of \emph{concavity-spoiling} breakpoints, i.e., the ones in which the arc-delay slopes increase. Clearly, $K^* \leq K$, and $K^* = 0$ for \emph{concave} pwl functions. The space to represent the entire network is $\Order{n+m+K}$.

The \emph{arc-arrival} function $Arr[uv](t_u) = t_u + D[uv](t_u)$ represents arrival-times at $v$, depending on the departure-times $t_u$ from $u$.
Note that we can express the same delay function of an arc $a=uv$ as a function of the arrival-time $t_v = t_u + D[uv](t_u)$ at the head $v$.
This is specifically useful when we need to work with the \emph{reverse network} $(\rev{G} = (V,A,(\rev{D}[a])_{a\in A})$,
where $\rev{D}[uv]$ is the delay of arc $a=uv$, measured now as a function of the arrival-time $t_v$ at $v$.
For instance, consider an arc $a=uv$ with $D[uv](t_u) = t_u + 1$, $0\le t_u \le 3$. Then, $t_v = 2t_u + 1$ and $1 \le t_v \le 7$.
Now, the same delay function can be expressed as a function of $t_v$ as $\rev{D}[uv](t_v) = t_v - t_u = t_v - \frac{t_v - 1}{2} = \frac{t_v + 1}{2}$,
for $1 \le t_v \le 7$.

For any $(o,d)\in V\times V$, $\mathcal{P}_{o,d}$ is the set of $od-$paths, and $\mathcal{P} = \union_{(o,d)} \mathcal{P}_{o,d}$. For a path $p\in\mathcal{P}$, $\subpath{p}{x}{y}$ is its subpath from (the  first appearance of) vertex $x$ until (the subsequent first appearance of) vertex $y$. For any pair of paths $p\in \mathcal{P}_{o,v}$ and $q\in \mathcal{P}_{v,d}$, $p\bullet q$ is the $od-$path produced as the concatenation of $p$ and $q$ at $v$.

For any path (represented as a sequence of arcs) $p = \langle a_1,a_2,\cdots,a_k\rangle\in \mathcal{P}_{o,d}$, the \emph{path-arrival} function is the composition of the constituent arc-arrival functions: $\forall t_o\in[0,T)$, $Arr[p](t_o) = Arr[a_k](Arr[a_{k-1}](\cdots(Arr[a_1](t_o))\cdots))$. The \emph{path-travel-time} function is $D[p](t_o) = Arr[p](t_o) - t_o$. The \emph{earliest-arrival-time} and \emph{shortest-travel-time} functions from $o$ to $d$ are: $\forall t_o\in[0,T), Arr[o,d](t_o) = \min_{p\in \mathcal{P}_{o,d}} \left\{ Arr[p](t_o) \right\}$ and $D[o,d](t_o) = Arr[o,d](t_o) - t_o$. Finally, $SP[o,d](t_o)$ (resp. $ASP[o,d](t_o)$) is the set of shortest (resp., with stretch-factor at most $(1+\eps)$) $od-$paths for a given departure-time $t_o$.

\subsection{Facts of the FIFO property}
\label{section:FIFO-facts}

We consider networks $(G=(V,A),(D[a])_{a\in A})$ with continuous arc-delay functions, possessing the \emph{FIFO} (a.k.a. \emph{non-overtaking}) property, according to which all arc-arrival-time functions are non-decreasing:
\begin{equation}
\label{eq:FIFO}
	\forall t_u, t'_u\in\reals, \forall uv\in A,
	t_u > t'_u \Rightarrow Arr[uv](t_u) \geq Arr[uv](t'_u)
\end{equation}
The FIFO property is \emph{strict}, if the above inequality is strict. The following properties (Lemmata~\ref{lemma:FIFO-characterization}--\ref{lemma:subpath-opt}),
are, perhaps, more-or-less known. We state them here and provide their proofs only for the sake of completeness.

\begin{lemma}[FIFO Property and Arc-Delay Slopes]
\label{lemma:FIFO-characterization}
	If the network satisfies the (strict) FIFO property then any arc-delay function has left and right derivatives with values at least (greater than) $-1$.
\end{lemma}
\begin{proof}
Observe that, by the FIFO property: $\forall a\in A, \forall t_u\in\reals,\forall \de>0$,
	\begin{eqnarray*}
	&&	Arr[a](t_u) \leq Arr[a](t_u+\de) \Leftrightarrow
		t_u + D[a](t_u) \leq t_u + \de + D[a](t_u+\de)
	\\
	& \DueTo{\Leftrightarrow}{\de>0} &
		\frac{D[a](t_u+\de) - D[a](t_u)}{\de} \geq -1
	\end{eqnarray*}
This immediately implies that the left and right derivatives of $D[a]$ are lower bounded (strictly, in case of strict FIFO property) by $-1$.
\qed
\end{proof}

It is easy to verify that the FIFO property also holds for arbitrary path-arrival-time functions
and earliest-arrival-time functions.

\begin{lemma}[FIFO Property for Paths]
\label{lemma:path-FIFO}
	If the network satisfies the FIFO property, then
\(
	\forall p\in\mathcal{P}, \forall t_1\in\reals, \forall \de>0,~ Arr[p](t_1) \leq Arr[p](t_1+\de)\,.
\)
In case of strict FIFO property, the inequality is also strict. The (strict) monotonicity holds also for $Arr[o,d]$.
\end{lemma}
\begin{proof}
To prove the FIFO property for a path $p=\langle a_1,\ldots,a_k\rangle\in\mathcal{P}$, we use a simple inductive argument on the prefixes of $p$, based on a recursive definition of path-arrival-time functions. $\forall 1\leq i \leq j \leq k$, let $p_{i,j}$ be the subpath of $p$ starting with the $i^{\rm th}$ arc $a_i$ and ending with the $j^{\rm th}$ arc $a_j$ in order. Then:
\begin{eqnarray}
\label{eq:path-arrival-recursive}
	\nonumber
	\lefteqn{Arr[p_{1,k}](t_o)}
\\ \nonumber
	&=&	t_o + D[p_{1,k}](t_o)
	   =		\underbrace{t_o + D[p_{1,1}](t_o)}_{= Arr[p_{1,1}](t_o)} + D[p_{2,k}](t_o + D[p_{1,1}](t_o))
	\\ \nonumber
	&=&	Arr[p_{2,k}]\left( Arr[p_{1,1}](t_o)\right) = \left( Arr[p_{2,k}]\circ Arr[p_{1,1}]\right)(t_o) = \cdots
	\\
	&=&	\left( Arr[a_k] \circ \cdots\circ Arr[a_1] \right)(t_o)
\end{eqnarray}
The composition of non-decreasing (increasing) functions is well known to also be non-decreasing (increasing).
Applying a minimization operation to produce the earliest-arrival-time function $Arr[o,d] = \min_{p\in \mathcal{P}_{o,d}} \left\{Arr[p]\right\}$, preserves the same kind of monotonicity.
\qed
\end{proof}
It is well-known that in FIFO (or equivalently, non-FIFO with unrestricted-waiting-at-nodes) networks the crucial property of \emph{prefix-subpath optimality} is preserved~\cite{Dreyfus1969}. We strengthen this observation to the more general (arbitrary) \emph{subpath optimality}, for strict FIFO networks.
\begin{lemma}[Subpath Optimality in strict FIFO Networks]
\label{lemma:subpath-opt}
	If the network possesses the strict FIFO property, then $\forall (u,v)\in V\times V$, $\forall t_u\in\reals$ and any optimal path
\(
	p^*\in SP[u,v](t_u)\,,
\)
it holds for every subpath $q^*\in \mathcal{P}_{x,y}$ of $p^*$ that $q^*\in SP[x,y](Arr[\subpath{p^*}{u}{x}](t_u))$. In other words, $q^*$ is a shortest path between its endpoints $x,y$ for the earliest-departure-time from $x$, given $t_u$.
\end{lemma}
\begin{proof}
Let $t^*_x = Arr[\subpath{p^*}{u}{x}](t_u)$. For sake of contradiction, assume that
\(
	\exists q \in \mathcal{P}_{x,y}: D[q](t^*_x) <  D[q^*](t^*_x)\,.
\)
Then, $p = \subpath{p^*}{u}{x} \bullet q\bullet \subpath{p^*}{y}{v}$ suffers smaller delay than $p^*$ for departure time $t_u$. Indeed, let $t_y \equiv  t^*_x + D[q]( t^*_x)$ and $ t^*_y\equiv  t^*_x +  D[\subpath{p^*}{x}{y}]( t^*_x)$. Due to the alleged suboptimality of $\subpath{p^*}{x}{y}$ when departing at time $ t^*_x$, it holds that $t_y <  t^*_y$. Then:
\begin{eqnarray*}
	Arr[p](t_u) &=& t_u + D[p](t_u)
	\\
	&=& \underbrace{t_u +  D[\subpath{p^*}{u}{x}](t_u)}_{= t^*_x} + D[q]( t^*_x) + D[\subpath{p^*}{y}{v}]( t^*_x+D[q]( t^*_x))
	\\
	&=& \underbrace{ t^*_x +  D[q]( t^*_x)}_{=t_y} + D[\subpath{p^*}{y}{v}]( t^*_x+D[q]( t^*_x))
	   =    t_y + D[\subpath{p^*}{y}{v}](t_y)
	\\
	&<&  t^*_y + D[\subpath{p^*}{y}{v}]( t^*_y) = Arr[p^*](t_u)
\end{eqnarray*}
violating the optimality of $p^*$ for the given departure-time $t_u$ (the inequality is due to the \emph{strict} FIFO property of the suffix-subpath $\subpath{p^*}{y}{v}$).
\qed
\end{proof}

\noindent
Lemma~\ref{lemma:subpath-opt} implies that both Dijkstra's label setting algorithm and Bellman-Ford label-correcting algorithm also work in time-dependent strict FIFO networks, under the usual conventions for static instances (positivity of arc-delays for Dijkstra, and inexistence of negative-travel-time cycles for Bellman-Ford).

In the following, we shall refer to an execution of the time-dependent Dijkstra's algorithm ($\alg{TDD}$) from origin $o\in V$,
with departure time $t_o\in[0,T)$, either as \emph{``a run of $\alg{TDD}$ from  $(o,t_o)$"}, or as \emph{``growing a $(\alg{TDD})$ ball around (or centered at) $(o,t_o)$ (by running $\alg{TDD}$)"}.

The time-complexity of $\alg{TDD}$ is slightly worse than the corresponding complexity of Dijkstra's algorithm in the static case, since during the relaxation of each arc the actual arc-travel-time of the arc has to be \emph{evaluated} rather than simply retrieved. For example, if each arc-travel-time function $D[uv]$ is periodic, continuous and pwl, represented by at most $K_{\max}$ breakpoints, then the evaluation of arc-travel-times can be done in time $\Order{\log\log(K_{\max})}$, e.g. by using a predecessor-search structure to determine the right leg for each function. Therefore, the time complexity for $\alg{TDD}$ would be $\Order{[m+n\log(n)]\log\log(K_{\max})}$.

\subsection{Towards a time-dependent distance oracle}
\label{section:Assumptions}

Our approach for providing a time-dependent distance oracle is inspired by the generic approach for general \emph{undirected} graphs under \emph{static} travel-time metrics. However, we have to tackle the two main challenges of \emph{directedness} and \emph{time-dependence}. Notice that together these two challenges imply an \emph{asymmetric} distance metric, which also \emph{evolves} with time. Consequently, to achieve a smooth transition from the static and undirected world towards the time-dependent and directed world, we have to quantify the \emph{degrees of asymmetry and evolution} in our metric.

Towards this direction, we introduce some metric-related parameters which quantify (i) the steepness of the shortest-travel-time functions (via the parameters $\La_{\min}$ and $\La_{\max})$, and (ii)
the degree of asymmetry (via the parameter $\zeta$).
We make two assumptions on the values of these parameters, namely, that they have constant (in particular, independent of the network size) values. These assumptions seem quite natural in realistic time-dependent route planning instances, such as urban-traffic metropolitan road networks.
The first assumption, called \emph{Bounded Travel-Time Slopes}, asserts that the partial derivatives of the shortest-travel-time functions between any pair of origin-destination vertices are bounded in a given fixed interval $[\La_{\min},\La_{\max}]$.
%
\begin{assumption}[Bounded Travel-Time Slopes]
\label{assumption:Bounded-Travel-Time-Slopes}
There are \emph{constants} $\La_{\min}\in[0,1)$ and $\La_{\max}\geq 0$ s.t.:
\(
	\forall (o,d)\in V\times V,~ \forall t_1 < t_2,~
	\frac{D[o,d](t_1) - D[o,d](t_2)}{t_1 - t_2} \in[-\La_{\min}, \La_{\max}]\,.
\)
\end{assumption}
%

The lower bound $-\La_{\min} > -1$ is justified by the FIFO property (cf. Lemmata~\ref{lemma:FIFO-characterization} and \ref{lemma:path-FIFO} in  Section~\ref{section:FIFO-facts}).
$\La_{\max}$ represents the maximum possible rate of change of shortest-travel-times in the network, which only makes sense to be bounded (in particular, independent of the network size) in realistic instances such as the ones representing urban-traffic time-dependent road networks.

Towards justifying this assumption, we conducted an experimental analysis with two distinct data sets. The first one is a real-world time-dependent snapshot of two weeks traffic data of the city of Berlin, kindly provided to us by TomTom \cite{ecompass} (consisting of $n = 478,989$ vertices and $m = 1,134,489$ arcs), in which the arc-delay functions are the continuous, pwl interpolants of five-minute samples of the average travel-times in each road segment. The second data set is a benchmark time-dependent instance of Western Europe's (WE) road network (consisting of $n = 18,010,173$ vertices and $m = 42,188,664$ arcs) kindly provided by PTV AG for scientific use. The time-dependent arc travel time functions were generated as described in \cite{ndls-bastd-12}, reflecting a high amount of traffic for all types of roads (highways, national roads,
urban roads), all of which posses non-constant time-dependent arc travel time functions.

We conducted $10000$ random queries $(o,d,t_o)$ in the Berlin (real-world) instance, focusing on the harder case of rush-hour departure times. We computed the approximate distance functions towards all the destinations using our one-to-all approximation algorithm (cf.~Section~\ref{section:preprocessing}) with approximation guarantee $\eps=0.001$. Our observration was that all the \emph{shortest-travel-time} slopes were $\La_{\max}\leq 0.1843$. That is, the travel-time on every road segment increases at a rate of at most $18,43\%$ as departure time changes. An analogous experimentation with the benchmark instance of Europe (heavy-traffic variant), again by conducting $10000$ random queries, demonstrated shortest-travel-time slopes at most $\La_{\max} \leq 6.1866$.

The second assumption, called \emph{Bounded Opposite Trips}, asserts that
for any given departure time, the shortest-travel-time from $o$ to $d$ is
not more than a  \emph{constant} $\zeta\geq 1$ times the shortest-travel-time
in the opposite direction (but not necessarily along the same path).
\begin{assumption}[Bounded Opposite Trips]
\label{assumption:Bounded-Opposite-Trips}
There is a constant $\zeta \geq 1$ such that:
\(
	\forall (o,d)\in V\times V,~ \forall t\in[0,T),~
	D[o,d](t) \leq \zeta\cdot D[d,o](t)\,.
\)
\end{assumption}
This is also a quite natural assumption in road networks, because it is most unlikely that a trip in one direction would be, say, more than $10$ times longer than the trip in the opposite direction (but not necessarily along the reverse path) during the same time period. This was also justified by the two instances at our disposal.
In each instance we uniformly selected $10000$ random origin-destination pairs and departure times randomly chosen from the rush-hour period,
which is the most interesting and diverging case. For the Berlin-instance the resulting worst-case value was $\zeta \leq 1.5382$.
For the WE-instance the resulting worst-case value was $\zeta \leq 1.174$.

\begin{figure}[h]
\begin{center}
\includegraphics[width=5.5cm, height=4cm]{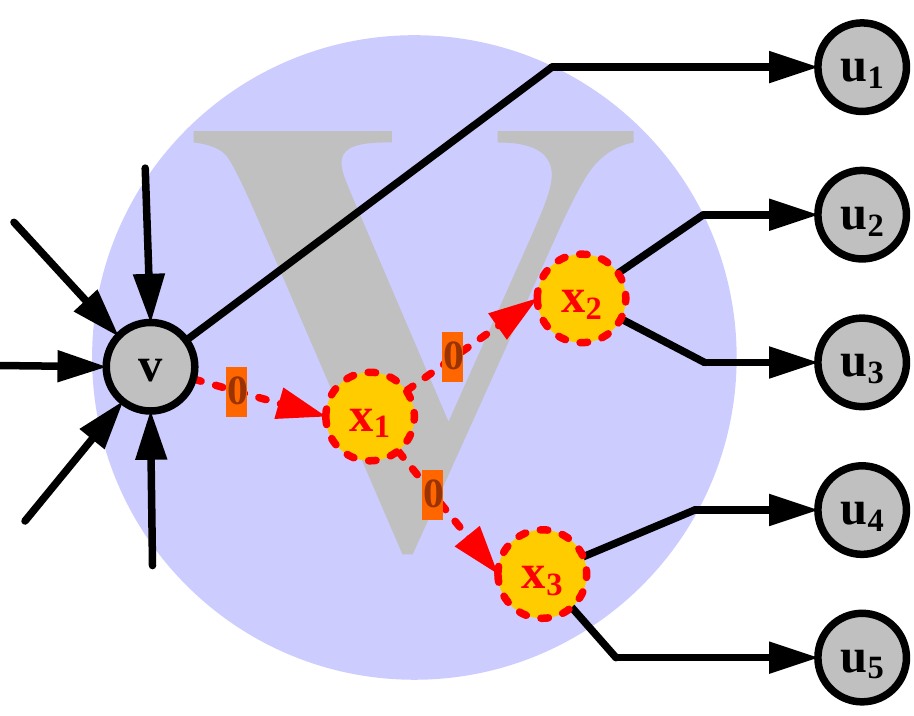}
\end{center}
\caption{\label{fig:KZ13_Bounded-Out-Degree-Construction} The node substitution operation for a vertex $v\in V$ with $d^{+}_G(v) = 5$. The operation ensures an out-degree at most $2$ for all the newly inserted vertices in place of $v$ in the graph. The new graph elements (nodes and arcs) are indicated by dashed (red) lines. The solid (black) arcs and vertices are the ones pre-existing in the graph.
	}
\end{figure}

A third assumption that we make is that the  maximum out-degree of every node is bounded by $2$. This can be easily guaranteed by using an equivalent network of at most double size (number of vertices and number of arcs). This is achieved by substituting every vertex of the original graph $(V,A)$ with out-degree greater than $2$ with a complete binary tree whose leaf-edges are the outgoing edges from $v$ in $(V,A)$, and each internal level consists of a maximal number of nodes with two children from the lower level, until a 1-node level is reached. This root node inherits all the incoming arcs from $v$ in the original graph. All the newly inserted arcs (except for the original arcs outgoing from $v$) get zero delay functions. Figure~\ref{fig:KZ13_Bounded-Out-Degree-Construction} demonstrates an example of such a substitution.
For each node $v\in G$ with out-degree $d^{+}(v) > 2$, the node substitution operation is executed in time $\Order{d^{+}(v)}$ and introduces $d^{+}(v) - 1$ new nodes and $d^{+}(v) - 2$ new arcs (of zero delays). Therefore, in time $\Order{|A|}$ we can ensure  out-degree at most $2$ and the same time-dependent travel-time characteristics, by at most doubling the size of the graph ($\sum_{v\in V: d^{+}(v) > 2} (d^{+}(v) - 1) < |A|$ new nodes and $\sum_{v\in V: d^{+}(v) > 2} (d^{+}(v) - 2) < |A|$ new arcs).

\subsection{Overview of our approach}
\label{section:Overview}
We follow (at a high level) the typical approach adopted for the construction of approximate distance
oracles in the static case. In particular, we start by selecting a subset $L\subset V$ of \term{landmarks},
i.e., vertices which will act as reference points for our distance summaries.
For our oracle to work, several ways to choose $L$ would be acceptable, that is,
we can choose the landmarks randomly among all vertices, or we can choose as landmarks the
vertices in the cut sets provided by some graph partitioning algorithm.
Nevertheless, for the sake of the analysis we assume that landmark selection is done
by deciding for each vertex randomly and independently with probability $\rho\in (0,1)$
whether it belongs to $L$. After having $L$ fixed, our approach is \emph{deterministic}.

We start by constructing (concurrently, per landmark) and storing the \term{distance summaries}, i.e., all landmark-to-vertex $(1+\eps)-$approximate travel-time functions, in time and space $\order{(1+K^*) n^2}$.
Then, we provide two approximation algorithms for responding to arbitrary queries $(o,d,t_o)\in V\times V\times [0,T)$.
The first ($\alg{FCA}$) is a simple \emph{sublinear}-time constant-approximation algorithm (cf. Section~\ref{section:constant-approx-alg}).
The second ($\alg{RQA}$) is a recursive algorithm growing small $\alg{TDD}$ outgoing balls from vertices in the vicinity of the origin, until either a satisfactory approximation guarantee is achieved, or an upper bound $r$ on the depth of the recursion (the \term{recursion budget}) has been exhausted. $\alg{RQA}$ finally responds with a $(1+\s)-$approximate travel-time to the query in \emph{sublinear} time, for any constant $\s > \eps$ (cf. Section~\ref{section:ptas-alg}).
As it is customary in the distance oracle literature, the query times of our algorithms concern the determination of (upper bounds on) shortest-travel-time from $o$ to $d$.
An actual path guaranteeing this bound can be reported in additional time that is linear in the number of its arcs
(cf.~Section~\ref{sec:path-reconstruction}).

\section{Preprocessing Distance Summaries}
\label{section:preprocessing}

The purpose of this section is to demonstrate how to construct the preprocessed
information that will comprise the \emph{distance summaries} of the oracle, i.e., all landmark-to-vertex $(1+\eps)$-approximate shortest-travel-time functions.

Our focus is on instances with \emph{concave}, continuous, pwl arc-delay functions possessing the strict FIFO property. If there exist $K^*\geq 1$ \emph{concavity-spoiling} breakpoints among the arc-delay functions, then we do the following:
For each of them (which is a departure-time $t_u$ from the tail $u$ of an arc $uv\in A$) we run a reverse variant of $\alg{TDD}$ (going ``back in time") with root $(u,t_u)$ on the \emph{network} $(\rev{G} = (V,A,(\rev{D}[a])_{a\in A})$, where $\rev{D}[uv]$ is the delay of arc $a=uv$, measured now as a function of the arrival-time $t_v$ at the head $v$. The algorithm proceeds backwards both along the connecting path (from the destination towards the origin) and in time. As a result, we compute all \emph{latest-departure-times} from landmarks that allow us to determine the images (i.e., projections to appropriate departure-times from all possible origins) of concavity-spoiling breakpoints to the spaces of departure-times from each of the landmarks.
Then, for each landmark, we repeat the procedure for concave, continuous, pwl arc-delay functions -- described in the rest of this section -- independently for each of the (at most) $K^*+1$ consecutive subintervals of $[0,T)$ determined by these consecutive images of concavity-spoiling breakpoints. Within each subinterval all arc-travel-time functions are concave, as required in our analysis.

We must construct in polynomial time, for all $(\ell,v)\in L\times V$, succinctly represented upper-bounding $(1+\eps)-$approximations $\De[\ell,v]: [0,T)\rightarrow \positivereals$ of the shortest-travel-time functions
$D[\ell,v]:[0,T)\rightarrow \positivereals$, i.e., for each $(\ell,v)\in L\times V$ we have to
compute a continuous pwl function $\De[\ell,v]$ with a \emph{constant} number of breakpoints, such that
\(
	\forall t_o\in[0,T),~ D[\ell,v](t_o)
	\leq \De[\ell,v](t_o)
	\leq (1+\eps)\cdot D[\ell,v](t_o)\,.
\)
%
An algorithm providing such functions in a \emph{point-to-point} fashion was proposed in \cite{Foschini-Hershberger-Suri-2011}. For each landmark $\ell\in L$, it has to be executed $n$ times so as to construct all the required landmark-to-vertex approximate functions. The main idea of that algorithm is to keep sampling the travel-time axis of the unknown function $D[\ell,v]$ at a logarithmically growing scale, until its slope becomes less than $1$. It then samples the departure-time axis via bisection, until the required approximation guarantee is achieved. All the sample points (in both phases) correspond to breakpoints of a lower-approximating function. The upper-approximating function has at most twice as many points. The number of breakpoints returned may be suboptimal, given the required approximation guarantee: even for an affine shortest-travel-time function with slope in $(1,2]$ it would require a number of points logarithmic in the ratio of max-to-min travel-time values from $\ell$ to $v$, despite the fact that we could avoid all intermediate breakpoints for the upper-approximating function.
\medskip

Our solution is an improvement of the approach in \cite{Foschini-Hershberger-Suri-2011} in three aspects:

\begin{itemize}

\item[(i)] It computes \emph{concurrently} all the required approximate distance functions from a given landmark, at a cost equal to that of a single (worst-case with respect to the given origin and all possible destinations) point-to-point approximation of \cite{Foschini-Hershberger-Suri-2011}.

\item[(ii)] Within every subinterval of consecutive images of concavity-spoiling breakpoints, it requires asymptotically optimal space per landmark, which is also independent of the network size per landmark-vertex pair, implying that the required preprocessing space per vertex is $\Order{|L|}$. This is also claimed in \cite{Foschini-Hershberger-Suri-2011}, but it is actually true only for their second phase (the bisection). For the first phase of their algorithm, there is no such guarantee.

\item[(iii)] It provides an exact closed form estimation (see below) of the worst-case absolute error, which guides our method.

\end{itemize}

In a nutshell, our approach constructs two continuous pwl-approximations of the unknown shortest-travel-time function $D[\ell,v]:[0,T)\rightarrow\positivereals$, an upper-bounding approximate function $\upperD[\ell,v]$ and a lower-bounding approximate function $\lowerD[\ell,v]$. $\upperD[\ell,v]$ plays the role of $\De[\ell,v]$. Our construction guarantees that the exact function is always ``sandwiched'' between these two approximations.

To achieve a concurrent one-to-all construction of upper-bounding approximations from a
given landmark $\ell\in L$, our algorithm is purely based on bisection. This is done because the departure-time axis is common for all these unknown functions $(D[\ell,v])_{v\in V}$. In order for this technique to work, despite the fact that the slopes may be greater than one, a crucial ingredient is an \emph{exact closed-form estimation} of the worst-case absolute error that we provide. This helps our construction to indeed consider only the necessary sampling points as breakpoints of the corresponding (concurrently constructed) shortest travel-time functions. It is mentioned that this guarantee could also be used in the first phase of the approximation algorithm in \cite{Foschini-Hershberger-Suri-2011}, in order to discard all unnecessary sampling points from being actual breakpoints in the approximate functions.
Consequently, we start by providing the closed form estimation of the maximum absolute error and then we present our one-to-all approximation algorithm.

\subsection{Absolute Error Estimation}
\label{section:Absolute-Error-Estimation}

In this section, we provide a closed form for the maximum absolute error between the upper-approximating and the lower-approximating functions of a generic shortest-travel-time function $D$ within a time interval $[t_s,t_f)\subseteq[0,T)$ that contains no other primitive image, apart possibly from its endpoints.

For an interval $[t_s,t_f)\subseteq[0,T)$, fix an unknown, but amenable to polynomial-time
sampling, continuous (not necessarily pwl) \emph{concave} function $D:[t_s,t_f)\rightarrow \positivereals$,
with right and left derivative values at the endpoints $\La^+(t_s), \La^{-}(t_f)$.
Assume that $\La^+(t_s) > \La^{-}(t_f)$ and $L = t_f - t_s > 0$.

Let $m = \frac{ D(t_f) - D(t_s) + t_s\cdot \La^{+}(t_s) - t_f\cdot \La^{-}(t_f) }{ \La^{+}(t_s) -\La^{-}(t_f) }$
and $\upperD_m = \La^+(t_s)\cdot (m-t_s) + D(t_s)$.
\begin{lemma}
\label{lemma:MAE}
For an interval $[t_s,t_f)\subseteq[0,T)$ and a concanve function $D:[t_s,t_f)\rightarrow \positivereals$
defined as above, consider the affine function $\lowerD$ passing via the points $(t_s,D(t_s)),~ (t_f,D(t_f))$.
Consider also the pwl function $\upperD$ with three breakpoints $(t_s,D(t_s)),~ (m,\upperD_m),~ (t_f,D(t_f))$.
Then, $\forall t\in [t_s,t_f), \lowerD(t) \leq D(t) \leq \upperD(t)$ and the \emph{maximum absolute error}
(MAE) between $\lowerD$ and $\upperD$ in $[t_s,t_f)$ is expressed by the following form:
\begin{equation*}
	MAE(t_s,t_f)
	= (\La^{+}(t_s) - \La^{-}(t_f))\cdot\frac{(m-t_s)\cdot(t_f-m)}{L} \leq \frac{L\cdot(\La^{+}(t_s) - \La^{-}(t_f))}{4}\,.
\end{equation*}
\end{lemma}
\begin{proof}
Consider the affine functions (see also Figure~\ref{fig:AbsoluteError}):
\[
\begin{array}{rcl}
	y(x) &=& \frac{D(t_f) - D(t_f)}{L}\cdot x + \frac{D(t_s) t_f - D(t_f) t_s}{L}\,,
	\\[5pt]
	y_s(x) &=& \La^+(t_s)\cdot (x-t_s) + D(t_s)\,,
	\\[5pt]
	y_f(x) &=& \La^{-}(t_f)\cdot (x-t_f) + D(t_f)\,.
\end{array}
\]
\begin{figure}[h]
\centerline{\includegraphics[width=\textwidth]{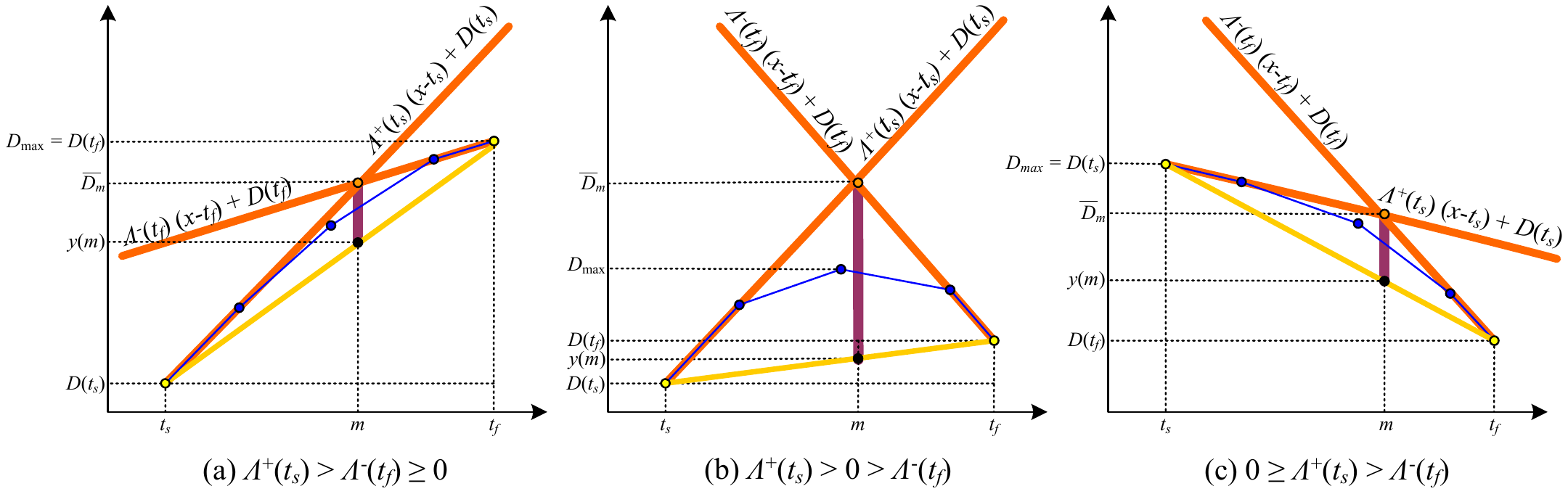}}
\caption{\label{fig:AbsoluteError}
	Three distinct cases for upper-bounding the absolute error between two consecutive interpolation points. The maximum absolute error (MAE) considered is shown by the vertical (purple) line segment at point $m$ of the time axis.
}
\end{figure}
The point $\left( m = \frac{ D(t_f) - D(t_s) + t_s\cdot \La^{+}(t_s) - t_f\cdot \La^{-}(t_f) }{ \La^{+}(t_s) -\La^{-}(t_f) }, \upperD_m = y_s(m) = y_f(m)\right)$ is the intersection point of the lines $y_s(x)$ and $y_f(x)$. As an upper-bounding (pwl) function of $D$ in $[t_s,t_f)$ we consider $\upperD(t) = \min\{y_s(t),y_f(t)\}$, whereas the lower-bounding (affine) function of $D$ is $\lowerD(t) = y(t)$.

By concavity and continuity of $D$, we know that the partial derivatives' values may only decrease with time, and at any given point in $[t_s,t_f)$ the left-derivative value is at least as large as the right-derivative value.
Thus, the restriction of $D$ on $[t_s,t_f)$ lies entirely in the area of the triangle $\{(t_s,D(t_s)),(m,\upperD_m),(t_f,D(t_f))\}$. The maximum possible distance (additive error) of $\upperD$ from $\lowerD$ is:
\[
	MAE(t_s,t_f) = \max_{t_s\leq t\leq t_f}\{ \upperD(t) - \lowerD(t)\}
\]
This value is at most equal to the \emph{vertical distance} of the two approximation functions, namely, at most equal to the length of the line segment connecting the points $(m,y(m))$ and $(m,\upperD_m)$(denoted by purple color in Figure~\ref{fig:AbsoluteError}). The calculations are identical for the three distinct cases shown in Figure ~\ref{fig:AbsoluteError}.
Let $\lowerLa = \frac{D(t_f) - D(t_s)}{L}$ be the slope of the line $y(x)$. Observe that:
\begin{eqnarray*}
	\lowerLa &=& \frac{D(t_f) - D(t_s)}{L} = \frac{(\upperD_m - D(t_s)) - (\upperD_m - D(t_f))}{L}
	\\
	&=& \frac{m - t_s}{L}\cdot\frac{\upperD_m - D(t_s)}{m - t_s} - \frac{t_f - m}{L}\cdot\frac{\upperD_m-D(t_f)}{t_f - m}
	\\
	&=& \frac{m - t_s}{L}\cdot \La^{+}(t_s) + \frac{t_f - m}{L}\cdot \La^{-}(t_f)\,.
\end{eqnarray*}
Thus we have:
\begin{eqnarray*}
	MAE(c,d)
	&=& 	\upperD_m - y(m) = (\upperD_m - D(t_s)) - (y(m) - D(t_s))
	\\
	&=& 	\La^{+}(t_s)\cdot(m - t_s) - \lowerLa\cdot(m - t_s) = (\La^{+}(t_s) - \lowerLa) \cdot (m - t_s)
	\\
	&=&	(\La^{+}(t_s) - \La^{-}(t_f))\cdot\frac{(m - t_s)\cdot(t_f - m)}{L}
	  \leq	\frac{L\cdot(\La^{+}(t_s) - \La^{-}(t_f))}{4}\,,
\end{eqnarray*}
since $(m - t_s) + (t_f - m) = t_f - t_s = L$ and the product $(m - t_s)\cdot(t_f - m)$ is maximized at $m = \frac{t_s + t_f}{2}$.
\qed
\end{proof}

\subsection{One-To-All Approximation Algorithm}

We now present our polynomial-time algorithm which provides asymptotically space-optimal succinct representations of one-to-all $(1+\eps)-$approximating functions $\bvec{D}[\ell,\star] = (\upperD[\ell,v])_{v\in V}$ of $D[\ell,\star] = (D[\ell,v])_{v\in V}$, for a given landmark $\ell\in L$ and all destinations $v\in V$, within a given time interval in which all the travel-time functions from $\ell$ are \emph{concave}. Recall our Assumption~\ref{assumption:Bounded-Travel-Time-Slopes} concerning the boundedness of the shortest-travel-time function slopes.  Given this assumption, we are able to construct a generalization of the bisection method proposed in \cite{Foschini-Hershberger-Suri-2011} for point-to-point approximations of distance functions, to the case of a single-origin $\ell$ and all reachable destinations from it.
Our method, which we call $\alg{BISECT}$, computes \emph{concurrently} (i.e., within the same bisection) all the required breakpoints to describe the (pwl) lower-approximating functions $\vector{\lowerD}[\ell,\star] = \left(\lowerD[\ell,v]\right)_{v\in V}$, and finally, via a linear scan of it, the upper-approximating functions $\vector{\upperD}[\ell,\star] = \left(\upperD[\ell,v]\right)_{v\in V}$.
This is possible because the bisection is done on the (common for all travel-time functions to approximate) axis of departure-times from the origin $\ell$.
The other crucial observation is that for each destination vertex $v\in V$ we keep as breakpoints of $\lowerD[\ell,v]$ only those sample points which are indeed necessary for the required approximation guarantee per particular vertex, thus achieving an asymptotically optimal space-complexity of our method, as we shall explain in the analysis of $\alg{BISECT}$. This is possible due to our closed-form expression for the (worst-case) approximation error between the lower-approximating and the upper-approximating distance function, per destination vertex  (cf. Lemma~\ref{lemma:MAE}). Moreover, all the travel-times from $\ell$ to be sampled at a particular bisection point $t_{\ell}\in [0,T)$ are calculated by a single time-dependent shortest-path-tree (e.g., $\alg{TDD}$) execution from $(\ell,t_{\ell})$.

Let the (unknown) concave travel-time function we wish to approximate be within a subinterval $[t_s,t_f)\subseteq [0,T)$.
Let $D_{\min}[\ell,v](t_s,t_f) = \min_{t\in [t_s,t_f]}\{D[\ell,v](t)\}$, and
$D_{\max}[\ell,v](t_s,t_f) = \max_{t\in [t_s,t_f]}\{D[\ell,v](t)\}$.
Due to the concavity of $D[\ell,v]$ in $[t_s,t_f]$, we have that $D_{\min}[\ell,v](t_s,t_f) = \min\{D[\ell,v](t_s),D[\ell,v](t_f)\}$.

The $\alg{BISECT}$ algorithm proceeds as follows:
for any subinterval $[t_s,t_f]\subseteq [0,T]$ we distinguish the destination vertices into \emph{active},
i.e., the ones for which the desired value $\eps\cdot D_{\min}[\ell,v](t_s,t_f)$ of the maximum absolute error
within $[t_s,t_f]$ (whose closed form is provided by Lemma~\ref{lemma:MAE}) has not been reached yet,
and the remaining \emph{inactive}.
Starting from $[t_s,t_f]=[0,T]$, as long as there is at least one active destination vertex for $[t_s,t_f]$, we bisect this time interval and recur on the subintervals $\left[t_s,(t_s+t_f)/2\right]$ and $\left[(t_s+t_f)/2 , t_f\right]$.
Prior to recurring to the two new subintervals, every destination vertex $v\in V$ that is active for $[t_s,t_f]$ stores the bisection point $(t_s+t_f)/2$ (and the corresponding sampled travel-time) in  a list $LBP[\ell,v]$ of breakpoints for $\lowerD[\ell,v]$. All inactive vertices just ignore this bisection point.
The bisection procedure is terminated as soon as all vertices have become inactive.

Apart from the list $LBP[\ell,v]$ of breakpoints for $\lowerD[\ell,v]$, a linear scan
of this list allows also the construction of the list $UBP[\ell,v]$ of breakpoints for $\upperD[\ell,v]$:
per consecutive pair of breakpoints in $LBP[\ell,v]$ that are added to $UBP[\ell,v]$, we must also
add their intermediate breakpoint $(m, \upperD_m)$ to $UBP[\ell,v]$ (cf.~proof of Lemma~\ref{lemma:MAE}).

In what follows, $L[\ell,v] = |LBP[\ell,v]|$ is the number of breakpoints for $\lowerD[\ell,v]$,
$U[\ell,v] = |UBP[\ell,v]|$ is the number of breakpoints for $\upperD[\ell,v]$ and,
finally, $U^*[\ell,v]$ is the minimum number of breakpoints of any $(1+\eps)-$upper
approximating function of $D[\ell,v]$, within the time-interval $[0,T)$.

The following theorem summarizes the space-complexity and time-complexity of our bisection method for providing concurrently one-to-all shortest-travel-time approximate travel-time functions in time-dependent instances with concave\footnote{If concavity is not ensured, then these numbers must be multiplied by $1+K^*$, since the proposed approximation procedure has to be repeated per subinterval of consecutive images of concavity-spoiling breakpoints.}, continuous, pwl arc-travel-time functions, with bounded shortest-travel-time slopes.
%
\begin{theorem}
\label{thm:SO-BISECTION-performance}
For a given $\ell\in L$ and any $v\in V$, $\alg{BISECT}$ computes an asymptotically optimal, independent of the network size, number of breakpoints
\\[3pt]
\centerline{%
\(	
	U[\ell,v] \leq 4 U^*[\ell,v] \leq 4\log_{1+\eps}\left(\frac{D_{\max}[\ell,v](0,T)}{D_{\min}[\ell,v](0,T)}\right)
	\in \Order{\frac{1}{\eps}\log\left(\frac{D_{\max}[\ell,v](0,T)}{D_{\min}[\ell,v](0,T)}\right)}
\)
}
\\[3pt]
where $D_{\max}[\ell,v](0,T)$ and $D_{\min}[\ell,v](0,T)$ denote the maximum and minimum shortest-travel-time values from $\ell$ to $v$ within $[0,T)$. The number $TDP$ of time-dependent (forward) shortest-path-tree probes for the construction of all the lists of breakpoints for $(\lowerD[\ell,v])_{v\in V}$, is:
\\[3pt]
\centerline{%
\(	
	TDP\in
	\Order{ 	
				\max_{v\in V}\left\{\log\left(\frac{T\cdot(\La_{\max}+1)}{\eps D_{\min}[\ell,v](0,T)}\right)\right\}
				\cdot
				\frac{1}{\eps}\cdot\max_{v\in V}\left\{\log\left(\frac{D_{\max}[\ell,v](0,T)}{D_{\min}[\ell,v](0,T)}\right)\right\}
	}
	\,.
\)
}
\end{theorem}
\begin{proof}
The time complexity of $\alg{BISECT}$ will be asymptotically equal to that of the worst-case point-to-point bisection from $\ell$ to some destination vertex $v$. In particular,  $\alg{BISECT}$ \emph{concurrently} computes the new breakpoints for the lower-bounding approximate distance functions of all the active nodes, within the same $\alg{TDD}$-run. This is because the departure-time axis is common for all the shortest-travel-time functions from the common origin $\ell$. Moreover, due to being able to (exactly) calculate the worst-case maximum absolute error per destination vertex in each interval of the bisection, the algorithm is able to deactivate (and thus, stop producing breakpoints for) those vertices which have already reached the required approximation guarantee. The already deactivated node will remain so until the end of the algorithm. Nevertheless, the bisection continues as long as there exists at least one active destination vertex.

We now bound the number of breakpoints produced by $\alg{BISECT}$. The initial departure-times interval to bisect is $[0,T)$. Assume that we are currently at an interval $[t_s,t_f)\subseteq[0,T)$, of length $t_f-t_s$. A new bisection halves this subinterval and creates new breakpoints at $\frac{t_s+t_f}{2}$, one for each vertex that remains active. Thus, at the $k-$th level of the recursion tree all the subintervals have length $L(k) = T / 2^k$. Since for any shortest-travel-time function and any subinterval $[t_s,t_f)$ of departure-times from $\ell$ it holds that $0 \leq \La^{+}[\ell,v](t_s) - \La^{-}[\ell,v](t_f) \leq \La_{\max}+1$ (cf. Assumption~\ref{assumption:Bounded-Travel-Time-Slopes}), the absolute error between $\lowerD[\ell,v]$ and $\upperD[\ell,v]$ in this interval is (by Lemma~\ref{lemma:MAE}) at most $\frac{L(k)\cdot (\La_{\max}+1)}{4}\leq \frac{ T \cdot (\La_{\max}+1) }{ 2^{k+2} }$. This implies that the bisection will certainly stop at a level $k_{\max}$ of the recursion tree at which for any subinterval $[t_s,t_f)\subseteq [0,T)$ and any destination vertex $v\in V$ the following holds:
\begin{eqnarray*}
	&& MAE[\ell,v](t_s,t_f) \leq \frac{T\cdot(\La_{\max}+1)}{2^{k_{\max}+2}}
	\leq \eps D_{\min}[\ell,v](t_s,t_f) \leq \eps D_{\min}[\ell,v](0,T)
\end{eqnarray*}
From this we conclude that setting
\begin{equation}
\label{eq:k-max}
	k_{\max} = \max_{v\in V}\left\{\ceil{\log_2\left(\frac{T\cdot(\La_{\max}+1)}{\eps D_{\min}[\ell,v](0,T)}\right)}\right\} - 2
\end{equation}
is a safe upper bound on the depth of the recursion tree.

On the other hand, the parents of the leaves in the recursion tree correspond to subintervals $[t_s,t_f)\subset [0,T)$ for which the absolute error of at least one vertex $v\in V$ is greater than $\eps D_{\min}[\ell,v](t_s,t_f)$, indicating that (in worst case) no pwl $(1+\eps)-$approximation may avoid placing at least one interpolation point in this subinterval. Therefore, the proposed bisection method $\alg{BISECT}$ produces at most twice as many interpolation points (to determine the lower-approximating vector function $\vector{\lowerD}[\ell,\star]$) required for any $(1+\eps)$-upper-approximation of $\vector{D}[\ell,\star]$. But, as suggested in \cite{Foschini-Hershberger-Suri-2011}, by taking as breakpoints the (at most two) intersections of the horizontal lines $(1+\eps)^j\cdot D_{\min}[\ell,v](0,T)$ with the (unknown) function $D[\ell,v]$, one would guarantee the following upper bound on the minimum number of breakpoints for any $(1+\eps)-$approximation of $D[\ell,v]$ within $[0,T)$:
\[
	U^*[\ell,v] \leq \ceil{\log_{1+\eps}\left( \frac{D_{\max}[\ell,v](0,T)}{D_{\min}[\ell,v](0,T)} \right)} - 1
\]
Therefore, $\forall v\in V$ it holds that:
\begin{equation}
\label{eq:L}
	L[\ell,v] \leq 2\cdot\log_{1+\eps}\left(\frac{D_{\max}[\ell,v](0,T)}{D_{\min}[\ell,v](0,T)}\right)
\end{equation}
The produced list $UBP[\ell,v]$ of breakpoints for the $(1+\eps)$-upper-approximation $\upperD[\ell,v]$ produced by $\alg{BISECT}$ uses at most one extra breakpoint for each pair of consecutive breakpoints in $LBP[\ell,v]$ for $\lowerD[\ell,v]$. Therefore, $\forall v\in V$:
\[
	U[\ell,v] \leq 4\cdot\log_{1+\eps}\left(\frac{D_{\max}[\ell,v](0,T)}{D_{\min}[\ell,v](0,T)}\right)
	\in \Order{\frac{1}{\eps}\log\left(\frac{D_{\max}[\ell,v](0,T)}{D_{\min}[\ell,v](0,T)}\right)}
\]

We now proceed with the time-complexity of $\alg{BISECT}$. We shall count the number $TDP$ of time-dependent shortest-path (TDSP) probes, e.g., $\alg{TDD}$ runs, to compute all the candidate breakpoints during the entire bisection. The crucial observation is that the bisection is applied on the common departure-time axis: In each recursive call from $[t_s,t_f]$, all the new breakpoints at the new departure-time $t_{mid} = \frac{t_s+t_f}{2}$, to be added to the breakpoint lists of the active vertices, are computed by a \emph{single} (forward) TDSP-probe. Moreover, for each vertex $v$, every breakpoint of $LBP[\ell,v](0,T)$ requires a number of (forward) TDSP-probes that is upper bounded by the path-length leading to the consideration of this point for bisection, in the recursion tree. Any root-to-node path in this tree has length at most $k_{\max}$, therefore each breakpoint of $LBP[\ell,v](0,T)$ requires at most $k_{\max}$ TDSP-probes, to be computed. In overall,
taking into account relations (\ref{eq:k-max}) and (\ref{eq:L}),
the total number $TDP$ of forward TDSP probes required to construct $LBP[\ell,\star](0,T)$, is upper-bounded by
\begin{eqnarray}
\lefteqn{TDP \leq k_{\max}\cdot\max_{v\in V}|LBP[\ell,v](0,T)|}
\nonumber
	\\	
\nonumber
	&\in&
	\Order{ 	
				\max_{v\in V}\left\{\ceil{\log\left(\frac{T (\La_{\max}+1)}{\eps D_{\min}[\ell,v](0,T)}\right)}\right\}
				\frac{1}{\eps} \max_{v\in V}\left\{\log\left(\frac{D_{\max}[\ell,v](0,T)}{D_{\min}[\ell,v](0,T)}\right)\right\}
	}
\end{eqnarray}
We can construct $UBP[\ell,\star](0,T)$ from  $LBP[\ell,\star](0,T)$ without any execution of a TDSP-probe, by just sweeping once for every vertex $v\in V$ $LBP[\ell,v](0,T)$ and adding all the intermediate breakpoints required. The time-complexity of this procedure is $\Order{|LBP[\ell,\star](0,T)|}$ and this is clearly dominated by the time-complexity (number of TDSP-probes) for constructing $LBP[\ell,\star](0,T)$ itself.
\qed
\end{proof}

Let $U = \max_{(\ell,v)\in L\times V}\{U[\ell,v]\}$. Theorem~\ref{thm:SO-BISECTION-performance} dictates that $U$ and $TDP$ are \emph{independent} of $n$ and they only depend on the degrees of asymmetry and time-dependence of the distance metric. Therefore, they can be treated as constants. Combining the performance of $\alg{BISECT}$ with the fact that the expected number of landmarks is $\Exp{|L|} = \rho n$, it is easy to deduce the required preprocessing time and space complexities for constructing all the $(1+\eps)-$approximate landmark-to-vertex distance summaries, which is culminated in the next theorem.

\begin{theorem}
\label{thm:preprocessing-complexity}
	The preprocessing phase of our time-dependent distance oracle has expected space/time complexities
	$\Exp{\mathcal{S}}\in$ $\Order{\rho n^2 (1+K^*) U}$
	and
	$\Exp{\mathcal{P}}\in$ $\Order{\rho n^2\log(n)\log\log(K_{\max}) (1+K^*) TDP}$.
\end{theorem}
\begin{proof}
For every landmark $\ell\in L$ and every destination vertex $d$, there are $(1+K^*)$
subintervals that need to be bisected and $\alg{BISECT}$ can generate at most $U$ new
breakpoints in each such interval. Since there are $n$ destinations, the total number
of breakpoints that need to be stored for all landmarks (and all destinations)
is $|L|n (1+K^*) U$. This total number of breakpoints can be computed concurrently
for all landmarks and all destinations (cf.~proof of Theorem~\ref{thm:SO-BISECTION-performance})
by $|L|(1+K^*) TDP$ runs of $\alg{TDD}$.
Each such runs takes time $\Order{(m+n\log(n))\log\log(K_{\max})}$, where
the extra $\log\log(K_{\max})$ term in the Dijkstra-time is due to the fact that
the arc-travel-times are continuous pwl functions of the departure-time from
their tails, represented as collections of breakpoints. A predecessor-search structure
would allow the evaluation of such a function to be achieved in time $\Order{\log\log(K_{\max})}$.
The space and time bounds now follow from the fact that $m=O(n)$ and $\Exp{|L|} = \rho n$.
\qed
\end{proof}

\section{Constant-approximation Query Algorithm}
\label{section:constant-approx-alg}

\begin{figure}[h]
\begin{center}
\includegraphics[width=7cm]{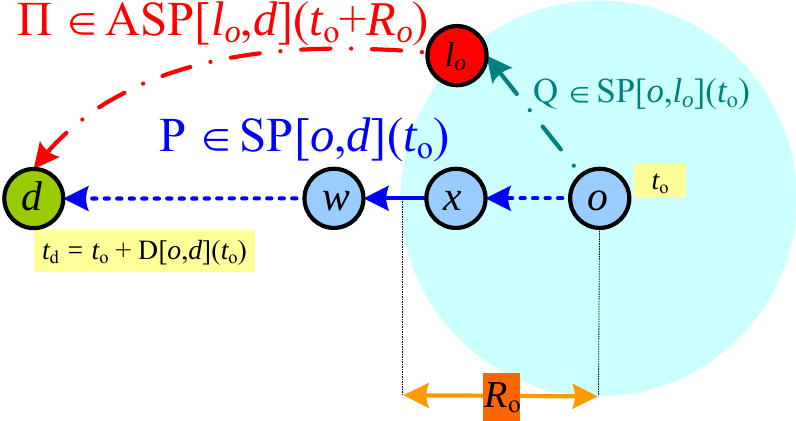}
\end{center}
\caption{\label{fig:KZ13_Constant-Approximation-Around-Origin}%
	The rationale of $\alg{FCA}$. The dashed (blue) path $P$ is a shortest $od-$path for $(o,d,t_o)$.
The dashed-dotted (green and red) path $Q\bullet \Pi$ is the via-landmark $od-$path indicated by the algorithm,
if the destination vertex is out of the origin's $\alg{TDD}$ ball.
}
\end{figure}

Our next step towards a distance oracle is to provide a fast query algorithm providing constant approximation to the shortest-travel-time values of arbitrary queries $(o,d,t_o)\in V\times V\times[0,T)$. The proposed query algorithm, called \term{Forward Constant Approximation} ($\alg{FCA}$), grows an outgoing ball
\[
	B_o := B[o](t_o) = \left\{x\in V : D[o,x](t_o) \leq \min\{ D[o,d](t_o) , D[o,\ell_o](t_o) \} \right\}
\]
from $(o,t_o)$, by running $\alg{TDD}$ until either $d$ or the closest landmark $\ell_o \in \arg\min_{\ell\in L}\{ D[o,\ell](t_o)\}$ is settled. We call $R_o = \min\{ D[o,d](t_o) , D[o,\ell_o](t_o) \}$ the \emph{radius} of $B_o$.
If $d\in B_o$, then $\alg{FCA}$ returns the \emph{exact} travel-time $D[o,d](t_o)$;
otherwise, it returns the approximate travel-time value $R_o + \De[\ell_o,d](t_o+R_o)$ via $\ell_o$.
Figure~\ref{fig:KZ13_Constant-Approximation-Around-Origin} gives an overview of the whole idea. Figure~\ref{fig:FCA-pseudocode} provides the pseudocode.

\begin{figure}[h]
\[
\begin{array}{|llr|}
\hline
\multicolumn{3}{|l|}{\alg{FCA}(o,d,t_o)}
\\ \hline\hline
1. & \IF o\in L ~\THEN \RETURN\left( \De[o,d](t_o) \right)
& \COMMENT{\mbox{\tiny $(1+\eps)-$approximate answer}}
\\
2. & \multicolumn{2}{l|}{B_o = \alg{TDD}\mbox{-ball around $(o,t_o)$ until either $d$ or the first landmark is settled}}
\\
3. & \IF d\in B_o ~\THEN \RETURN\left( D[o,d](t_o) \right)
&  \COMMENT{\mbox{\tiny exact answer}}
\\
4.	& \ell_o = B_o\intersection L; R_o = D[o,\ell_o](t_o);
& 
\\
5. & \RETURN\left( R_o + \De[\ell_o,d](t_o+R_o) \right)
& \COMMENT{\mbox{\tiny $(1+\eps+\psi)-$approximation}}
\\ \hline
\end{array}
\]
\caption{\label{fig:FCA-pseudocode}%
	The pseudocode describing $\alg{FCA}$.
}
\end{figure}

\subsection{Correctness of $\alg{FCA}$}

The next theorem demonstrates that $\alg{FCA}$ returns $od-$paths whose travel-times are constant approximations to the shortest travel-times.
\begin{theorem}
\label{theorem:Constant-Approximation-Around-Origin}
$\forall (o,d,t_o)\in V\times V\times [0,T)$, $\alg{FCA}$   returns either an exact path $P\in SP[o,d](t_o)$, or a via-landmark $od-$path $Q\bullet \Pi$, s.t. $Q\in SP[o,\ell_o](t_o)$, $\Pi\in ASP[\ell_o,d](t_o+R_o)$, and
\(
	D[o,d](t_o) 
	\leq R_o + \De[\ell_o,d](t_o + R_o) 
	\leq (1+\eps)\cdot D[o,d](t_o) + \psi\cdot R_o 
	\leq (1+\eps+\psi)\cdot D[o,d](t_o)\, ,
\)
where $\psi = 1 + \La_{\max}(1+\eps)(1+2\zeta+\La_{\max} \zeta) + (1+\eps)\zeta$.
\end{theorem}
\begin{proof}
In case that $d\in B_o$, there is nothing to prove since $\alg{FCA}$ returns the exact distance. So, assume that $d\notin B_o$, implying that $D[o,d](t_o) \geq R_o$. As for the returned distance value $R_o + \De[\ell_o,d](t_o + R_o)$, it is not hard to see that this is indeed an overestimation of the actual distance $D[o,d](t_o)$. This is because $\De[\ell_o,d](t_o + R_o)$ is an overestimation (implying also a connecting $\ell_od-$path) of $D[\ell_o,d](t_o + R_o)$, and of course $R_o = D[o,\ell_o](t_o)$ corresponds to a (shortest) $o\ell_o-$path that was discovered by the algorithm on the fly. Therefore, $R_o + \De[\ell_o,d](t_o + R_o)$ is an overestimation of an actual $od-$path for departure time $t_o$, and cannot be less than $D[o,d](t_o)$. We now prove that it is not arbitrarily larger than this shortest distance:
\begin{eqnarray*}
\lefteqn{R_o + \De[\ell_o,d](t_o + R_o) \leq R_o + (1+\eps) D[\ell_o,d](t_o + R_o)}
\\
&\DueTo{\leq}{\mbox{\tiny triangle}}&
R_o + (1+\eps)[D[\ell_o,o](t_o + R_o) + D[o,d](t_o + R_o + D[\ell_o,o](t_o + R_o))]
\\
&\DueTo{\leq}{\mbox{\tiny Assum.\ref{assumption:Bounded-Travel-Time-Slopes}}}&
R_o + (1+\eps)[(1+\La_{\max})D[\ell_o,o](t_o + R_o) + \La_{\max} R_o
                 + D[o,d](t_o)]
\\
&\DueTo{\leq}{\mbox{\tiny Assum.\ref{assumption:Bounded-Opposite-Trips}}}&
R_o + (1+\eps)[(1+\La_{\max})\zeta D[o,\ell_o](t_o + R_o) + \La_{\max} R_o
                 + D[o,d](t_o)]
\\
&\DueTo{\leq}{\mbox{\tiny Assum.\ref{assumption:Bounded-Travel-Time-Slopes}}}&
R_o + (1+\eps)[(1+\La_{\max}) \zeta (R_o + \La_{\max} R_o) + \La_{\max} R_o
                 + D[o,d](t_o)]
\\
&=&
\left[1 + (1+\eps)(1+\La_{\max})^2 \zeta + (1+\eps)\La_{\max}\right] R_0
+ (1+\eps) D[o,d](t_o)
\\
&=&
\underbrace{\left[1 + \La_{\max}(1+\eps)(1+2\zeta+\La_{\max} \zeta) + (1+\eps)\zeta\right]}_{=\psi} R_0
+ (1+\eps)  D[o,d](t_o)
\\
&=& (1+\eps)\cdot D[o,d](t_o) + \psi\cdot R_o
\end{eqnarray*}
{\flushright\qed}
\end{proof}

Note that $\alg{FCA}$ is a generalization of the $3-$approximation algorithm in \cite{Agarwal-Godfrey-2013} for symmetric (i.e., $\zeta = 1$) and time-independent (i.e., $\La_{\min} = \La_{\max} = 0$) network instances, the only difference being that the stored distance summaries we consider are $(1+\eps)-$approximations of the actual shortest-travel-times. Observe that our algorithm smoothly departs, through the parameters $\La_{\min}, \La_{\max}$ and $\zeta$, towards both \emph{asymmetry} and \emph{time-dependence} of the travel-time metric.

\subsection{Complexity of $\alg{FCA}$}
The main cost of $\alg{FCA}$ is to grow the ball $B_o = B[o](t_o)$ by running $\alg{TDD}$. Therefore, what really matters is the number of vertices in $B_o$, since the maximum out-degree is $2$. Recall that $L$ is chosen randomly by selecting each vertex $v$ to become a landmark independently of other vertices, with probability $\rho\in(0,1)$. Hence, for any $o\in V$ and any departure-time $t_o\in [0,T)$,
the size of the outgoing $\alg{TDD}-$ball $B_o = B[o](t_o)$ centered at $(o,t_o)$ until the first landmark vertex
is settled, behaves as a geometric random variable with success probability $\rho \in (0,1)$.
Consequently, $\Exp{|B_o|} = 1/\rho$, and moreover (as a geometrically distributed random variable), $\forall k\geq 1~, \Prob{|B_o| > k} = (1-\rho)^k \leq e^{-\rho k}$. By setting $k = (1/\rho)\ln(1/\rho)$ we conclude that: $\Prob{|B_o| > (1/\rho)\ln(1/\rho)} \leq  \rho$. Since the maximum out-degree is $2$, $\alg{TDD}$ will relax at most $2k$ arcs. Hence,
we have established the following.

\begin{theorem}
For the query-time complexity $\mathcal{Q}_{FCA}$ of $\alg{FCA}$ the following hold:
\begin{eqnarray*}
	&& \Exp{\mathcal{Q}_{FCA}} \in \Order{(1/\rho)\ln(1 / \rho)\log\log(K_{\max})}\,.
	\\
	&& \Prob{\mathcal{Q}_{FCA} \in \OmegaOrder{(1/\rho)\ln^2(1/\rho)\log\log(K_{\max})}} \in \Order{\rho}\,.
\end{eqnarray*}
\end{theorem}

\section{$(1+\s)-$approximate Query Algorithm}
\label{section:ptas-alg}

The \term{Recursive Query Algorithm} $\alg{(RQA)}$ improves the approximation guarantee of the chosen $od-$path
provided by $\alg{FCA}$, by exploiting carefully a number of recursive calls of $\alg{FCA}$,
based on a given bound -- called the \term{recursion budget} $r$ -- on the depth of the recursion tree to be constructed. Each of the recursive
calls accesses the preprocessed information and produces another candidate $od-$path.
The crux of our approach is the following: We ensure that, unless the required approximation guarantee has already been reached by a candidate solution, the recursion budget must be exhausted and the sequence of radii of the consecutive balls that we grow from centers lying on the unknown shortest path, is lower-bounded by a \emph{geometrically increasing} sequence. We prove that this sequence can only have a \emph{constant} number of elements until the required approximation guarantee is reached, since the sum of all these radii provides a lower bound on the shortest-travel-time that we seek.

A similar approach was proposed for \emph{undirected} and \emph{static} sparse networks~\cite{Agarwal-Godfrey-2013}, in which a number of recursively growing balls (up to the recursion budget) is used in the vicinities of \emph{both} the origin \emph{and} the destination nodes, before eventually  applying a constant-approximation algorithm to close the gap, so as to achieve improved approximation guarantees.

\begin{figure}[h]
\begin{center}
	\includegraphics[width=0.8\textwidth]{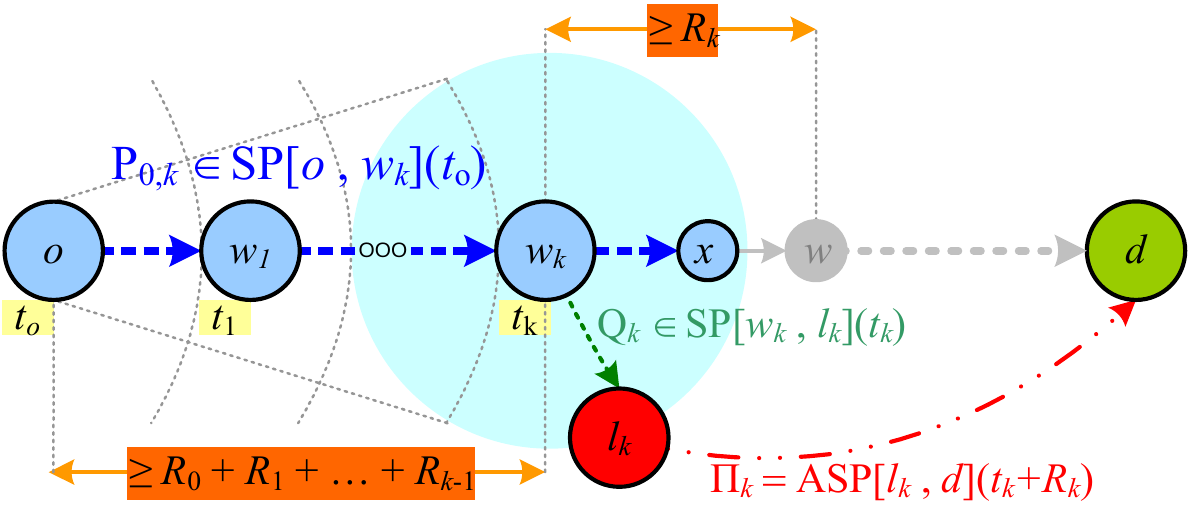}
\end{center}
	\caption{\label{fig:RQA-description} Overview of the execution of $\alg{RQA}$.}
	\end{figure}

In our case the network is both directed and time-dependent. Due to our ignorance of the exact arrival time at the destination, it is difficult (if at all possible) to grow incoming balls in the vicinity of the destination node. Hence, our only choice is to build a recursive argument that grows outgoing balls in the vicinity of the origin, since we only know the requested departure-time from it. This is exactly what we do: As long as we have not discovered the destination node within the explored area around the origin, and there is still some remaining recursion budget $r-k > 0$ ($k\in\{0,\ldots, r\}$), we ``guess'' (by exhaustively searching for it) the next node $w_k$ along the (unknown) shortest $od-$path. We then grow a new out-ball from the new center $(w_k, t_k = t_o + D[o,w_k](t_o))$, until we reach the closest landmark-vertex $\ell_k$ to it, at distance $R_k = D[w_k,\ell_k](t_k)$. This new landmark offers an alternative $od-$path $sol_k = P_{o,k}\bullet Q_k\bullet \Pi_k$ by a new application of $\alg{FCA}$, where $P_{o,k}\in SP[o,w_k](t_o),~ Q_k\in SP[w_k,\ell_k](t_k)$, and $\Pi_k\in ASP[\ell_k,d](t_k+R_k)$ is the approximate suffix subpath provided by the distance oracle. Observe that $sol_k$ uses a \emph{longer} optimal prefix-subpath $P_k$ which is then completed with a shorter approximate suffix-subpath $Q_k\bullet\Pi_k$.
Figure~\ref{fig:RQA-description} provides an overview of $\alg{RQA}$'s execution.
Figure~\ref{fig:RQA-pseudocode} provides the pseudocode of $\alg{RQA}$.

\begin{figure}[h]
\begin{small}
\[
\begin{array}{|ll|}
\hline
\multicolumn{2}{|l|}{\alg{RQA}(o,d,t_o,r)}
\\
\hline\hline
1. & \IF o \in L ~\THEN \RETURN\left(ASP[o,d](t_o), \De[o,d](t_o)\right)
		\hfill\COMMENT{\mbox{\tiny $(1+\eps)-$approximation}}
\\
2. & B[o](t_o) := \alg{TDD}\mbox{-ball from $(o,t_o)$ until either $d$ or a landmark is settled}
\\
3.	& \IF d\in B_o ~\THEN \RETURN\left(D[o,d](t_o)\right)
\hfill\COMMENT{exact suffix-subpath}
\\
4. & \ell_0\in B[o](t_o) \intersection L;~ R_0 = D[o,\ell_0](t_o)
\\
5. & sol_0 = \left(Q_0 \bullet \Pi_0 ~,~ \De[sol_0](t_o) = R_0 + \De[\ell_0,d](t_o+R_0)\right)
	\hfill\COMMENT{\mbox{\tiny via-$\ell_o$ approximation}}
\\
6. & k := 0;~ t_k = t_o;
\\
7. & \WHILE k < r \DO
\\
7.1. & \TAB \mbox{``guess'' the first vertex $w_{k+1}\in SP[w_k,d](t_k)\setminus B[w_k](t_k)$}
\hfill\COMMENT{\mbox{\tiny exhaustive search}}
\\
7.2. & \TAB t_{k+1} = t_k + D[w_k,w_{k+1}](t_k);
\\
7.3. & \TAB \IF w_{k+1}\in L
\\
7.4. & \TAB \THEN \RETURN\left( P_{0,k+1}\bullet \Pi[w_{k+1},d](t_{k+1}) , t_{k+1}-t_0 + \De[w_{k+1},d](t_{k+1}) \right)
\\
\multicolumn{2}{|r|}{\COMMENT{\mbox{\tiny approximate answer via $w_{k+1}$}}}
\\
7.5. & \TAB B[w_{k+1}](t_{k+1}) := \alg{TDD}\mbox{-ball until $d$ or a landmark is settled}
\\
7.6  & \TAB \IF d\in B[w_{k+1}](t_{k+1}) ~\THEN
\\
7.7. & \TAB \THEN sol_{k+1}
		= 	\left(\begin{array}{c}
				P_{0,k+1} \bullet P_{k+1,d},
				\\
				\De[sol_{k+1}](t_o) = t_{k+1} - t_o + D[w_{k+1},d](t_{k+1})
			\end{array}\right)
\\
7.8. & \TAB \ELSE
\\
7.8.1 & \TAB\TAB 	\ell_{k+1}\in L\intersection B[w_{k+1}](t_{k+1});~
						R_{k+1} = D[w_{k+1},\ell_{k+1}](t_{k+1})
\\
7.8.2 & \TAB\TAB 	sol_{k+1}
						= 	\left(\begin{array}{l}
								P_{0,k+1} \bullet Q_{k+1} \bullet \Pi_{k+1} ,
								\\[3pt]
								\De[sol_{k+1}](t_o) = t_{k+1} - t_o + R_{k+1}
								\\
								\TAB + \De[\ell_{k+1},d](t_{k+1}+R_{k+1})
						\end{array}\right)
\\
\multicolumn{2}{|r|}{\COMMENT{\mbox{\tiny approximate answer via $\ell_{k+1}$}}}
\\
7.9. & \TAB k = k+1
\\
8. & \END\WHILE
\\
9. & \RETURN \min_{0\leq k\leq r}\left\{ sol_k\right\}
\\ \hline
\end{array}
\]
\end{small}
\caption{\label{fig:RQA-pseudocode}%
	The recursive algorithm $\alg{RQA}$ providing $(1+\s)-$approximate time-dependent shortest paths. $Q_k\in SP[w_k,\ell_k](t_k)$ is the shortest path connecting $w_k$ to its closest landmark w.r.t. departure-time $t_k$. $P_{0,k}\in SP[o,w_k](t_o)$ is the prefix of the shortest $od-$path that has been already discovered, up to vertex $w_k$. $\Pi_k = ASP[\ell_k,d](t_k+R_k)$ denotes the $(1+\eps)-$approximate shortest $\ell_k d-$path precomputed by the oracle.}
\end{figure}

\subsection{Correctness and Quality of $\alg{RQA}$}
\label{subsection:RQA:correctness-quality}

The correctness of $\alg{RQA}$ implies that the algorithm always returns some $od-$path. This is true due to the fact that it either discovers the destination node $d$ as it explores new nodes in the vicinity of the origin node $o$, or it returns the shortest of the approximate $od-$paths $sol_0,\ldots,sol_r$ via one of the closest landmarks $\ell_o,\ldots,\ell_r$ to ``guessed'' nodes $w_0 = o, w_1,\ldots,w_r$ along the shortest $od-$path $P \in SP[o,d](t_o)$, where $r$ is the recursion budget. Since the preprocessed distance summaries stored by the oracle provide approximate travel-times corresponding to actual paths from landmarks to vertices in the graph, it is clear that $\alg{RQA}$ always returns an $od-$path whose travel-time does not exceed the alleged upper bound on the actual distance.

Our next task is to study the quality of the $1+\s$ stretch provided by $\alg{RQA}$. Let $\de>0$ be a parameter such that $\s = \eps+\de$ and recall the definition of $\psi$ from Theorem~\ref{theorem:Constant-Approximation-Around-Origin}. The next lemma shows that the sequence of ball radii grown from vertices of the shortest $od-$path $P[o,d](t_o)$ by the recursive calls of $\alg{RQA}$ is lower-bounded by a geometrically increasing sequence.
\begin{lemma}
\label{lemma:radii-geometric-increase}
Let $D[o,d](t_o) = t_d - t_o$ and suppose that $\alg{RQA}$ does not discover $d$ or any
landmark $w_k\in SP[o,d](t_o)\intersection L$, $k\in \{0,1,\ldots,r\}$, in the explored
area around $o$. Then, the entire recursion budget $r$ will be consumed and  in each
round $k$ of recursively constructed balls we have that either
\(
	R_k  >  \left(1+\frac{\eps}{\psi}\right)^k \cdot \frac{\de}{\psi}\cdot(t_d - t_o)
\)
or
\(
	\exists~ i\in\{0,1,\ldots,k\} :  D[sol_i](t_o) \leq (1+\eps+\de)\cdot D[o,d](t_o)\,.
\)
\end{lemma}
\begin{proof}
As long as none of the discovered nodes $o=w_0,w_1,\ldots,w_k$ is a landmark node and the recursion budget has not been consumed yet, $\alg{RQA}$ continues guessing new nodes of $P\in SP[o,d](t_o)$. If any of these nodes (say, $w_k$) is a landmark node, the $(1+\eps)-$approximate solution $P_{0,k}\bullet \Pi[w_k,d](t_k)$ is returned and we are done. Otherwise, $\alg{RQA}$ will certainly have to consume the entire recursion budget.

For any $k\in\{0,1,\ldots,r\}$, if $\exists~ i\in\{0,1,\ldots,k\}: D[sol_i](t_o) \leq (1+\eps+\de)\cdot D[o,d](t_o)$ then there is nothing to prove from that point on. The required disjunction trivially holds for all rounds $k, k+1,\ldots,r$. We therefore consider the case in which up to round $k-1$ of the recursion no good approximation has been discovered, and we shall prove inductively that either $sol_k$ is a $(1+\eps+\de)-$ approximation, or else
$R_k  >  \left(1+\frac{\eps}{\psi}\right)^k \cdot \frac{\de}{\psi}\cdot(t_d - t_o)$.

\noindent
\emph{Basis.}  Recall that $\alg{FCA}$ is used to provide the suffix-subpath of the returned solution $sol_0$, whose prefix (from $o$ to $\ell_o$) is indeed a shortest path. Therefore:
\[
\begin{array}{rl}
	\lefteqn{D[sol_0](t_o) \leq
	R_0 + \De[\ell_0,d](t_o+R_0)}
	\\
	& \DueTo{\leq} {\mbox{\tiny Theorem~\ref{theorem:Constant-Approximation-Around-Origin}}}
	(1+\eps)\cdot D[o,d](t_o) + \psi\cdot R_0
	=	\left( 1+\eps+\frac{\psi R_0}{t_d-t_o}\right)\cdot (t_d - t_o)
\end{array}
\]
Clearly, either $\frac{\psi R_0}{t_d-t_o}\leq \de \Leftrightarrow R_0 \leq \frac{\de}{\psi}\cdot (t_d-t_o)$, which then implies that we already have a $(1+\eps+\de)-$approximate solution, or else it holds that  $R_0 > \frac{\de}{\psi}\cdot (t_d-t_o)$.

\noindent
\emph{Hypothesis.} We assume inductively that $\forall~ 0\leq i\leq k$, no $(1+\eps+\de)-$approximate solution has been discovered up to round $k$, and thus it holds that  $R_i > \left(1+\frac{\de}{\psi}\right)^i\cdot \frac{\de}{\psi}\cdot (t_d-t_o)$.

\noindent
\emph{Step.} We prove that for the $(k+1)-$st recursive call, either the new via-landmark
solution $sol_{k+1} = P_{0,k+1}\bullet Q_{k+1} \bullet \Pi_{k+1}$ is a $(1+\eps+\de)-$approximate solution, or else  $R_{k+1} > \left(1+\frac{\de}{\psi}\right)^{k+1}\cdot \frac{\de}{\psi}\cdot (t_d-t_o)$. For the travel-time along this path we have:
\begin{eqnarray*}
\lefteqn{\TAB D[sol_{k+1}](t_o)}
\\
	&\leq& t_{k+1} - t_o + R_{k+1} + \De[\ell_{k+1},d](t_{k+1}+R_{k+1})
\\
	& \DueTo{\leq}{\mbox{\tiny Theorem~\ref{theorem:Constant-Approximation-Around-Origin}}} &
	t_{k+1} - t_o + (1+\eps)\cdot D[w_{k+1},d](t_{k+1}) + \psi\cdot R_{k+1}
\\
	&\DueTo{=}{w_{k+1}\in SP[o,d](t_o)}& t_{k+1} - t_o + (1+\eps)\cdot (t_d-t_{k+1}) + \psi\cdot R_{k+1}
\\
	&=& (1+\eps)\cdot (t_d-t_o) - \eps\cdot(t_{k+1}-t_o)+ \psi\cdot R_{k+1}
\\
	&\DueTo{\leq}{t_{k+1}-t_o \geq R_0+\ldots+R_k}& (1+\eps)\cdot (t_d-t_o) - \eps\cdot(R_0+\ldots+R_{k})+ \psi\cdot R_{k+1}
\\
	&\DueTo{<}{\mbox{\tiny Induction Hypothesis}}& (1+\eps)(t_d-t_o) - \eps \sum_{i=0}^k \left(1+\frac{\eps}{\psi}\right)^i \frac{\de}{\psi} (t_d-t_o) + \psi R_{k+1}
\end{eqnarray*}
\begin{eqnarray*}
	&=& \left(1+\eps - \frac{\eps\de}{\psi} \cdot \sum_{i=0}^k \left(1+\frac{\eps}{\psi}\right)^i + \frac{\psi\cdot R_{k+1}}{t_d-t_o}\right)\cdot (t_d-t_o)
\\
	&=& \left(1+\eps - \de\cdot \left[\left(1+\frac{\eps}{\psi}\right)^{k+1} - 1\right] + \frac{\psi\cdot R_{k+1}}{t_d-t_o}\right)\cdot (t_d-t_o)
\end{eqnarray*}
Once more, it is clear that either $D[sol_{k+1}](t_o)\leq (1+\eps+\de)\cdot D[o,d](t_o)$, or else it must hold that  $R_{k+1}>\left(1+\frac{\eps}{\psi}\right)^{k+1}\cdot \frac{\de}{\psi}\cdot(t_d-t_o)$ as required.
\qed
\end{proof}

The next theorem shows that $\alg{RQA}$ indeed provides $(1+\s)-$approximate distances in response to arbitrary queries $(o,d,t_o)\in V\times V\times [0,T)$.
\begin{theorem}
\label{thm:RQA-approximation-guarantee}
For the stretch of $\alg{RQA}$ the following hold:
\begin{enumerate}

	\item If $r = \ceil{\frac{\ln\left(1+\frac{\eps}{\de}\right)}{\ln\left(1+\frac{\eps}{\psi}\right)}} - 1$ for $\de>0$, then, $\alg{RQA}$ guarantees a stretch $1+\s = 1+\eps + \de$.

	\item For a given recursion budget $r\in \naturals$, $\alg{RQA}$ guarantees stretch $1+\s$,
where $\s = \s(r) \leq \eps\cdot\frac{(1+\eps/\psi)^{r+1}}{(1+\eps/\psi)^{r+1}-1}$.
\end{enumerate}
\end{theorem}
\begin{proof}
If none of the via-landmark solutions is a $(1+\eps+\de)-$approximation, then:
\begin{eqnarray*}
t_d - t_o &\geq& R_0 + R_1 + \ldots + R_r
\DueTo{>}{\mbox{\tiny Lemma~\ref{lemma:radii-geometric-increase}}} \frac{\de}{\psi}\cdot (t_d-t_o)\cdot\sum_{i=0}^r \left(1+\frac{\eps}{\psi}\right)^i
\\
&=&  \frac{\de}{\psi}\cdot (t_d-t_o)\cdot \frac{\left(1+\frac{\eps}{\psi}\right)^{r+1} - 1}{1+\frac{\eps}{\psi} - 1}
= \frac{\de}{\eps}\cdot (t_d-t_o)\cdot \left[\left(1+\frac{\eps}{\psi}\right)^{r+1} - 1\right]
\\
\Rightarrow \frac{\eps}{\de} &>& \left(1+\frac{\eps}{\psi}\right)^{r+1} - 1
\Rightarrow
	\left\{
	\begin{array}{rcl}
	r 	& < & \frac{\ln\left(1 + \eps/\de\right)}{\ln\left(1 + \eps/\psi\right)} - 1
	\\[10pt]
	\de	& < & \frac{\eps}{(1+\eps/\psi)^{r+1} - 1}
	\end{array}
	\right.
\end{eqnarray*}
If $r = \ceil{\frac{\ln\left(1 + \eps/\de\right)}{\ln\left(1 + \eps/\psi\right)} - 1} \leq \frac{\psi / \de}{1 - \eps/\psi} - 1\in \Order{\frac{\psi}{\de}}$, we have reached a contradiction\footnote{The inequality $r \leq \frac{\psi / \de}{1 - \eps/\psi} - 1$ holds due to the following bound: $\forall z\geq -\frac{1}{2}, z-z^2 \leq \ln(1+z)\leq z$.}. For this value of the recursion budget $\alg{RQA}$ either discovers the destination node, or at least a landmark node that also belongs to $SP[o,d](t_o)$, or else it returns a via-landmark path that is a $(1+\eps+\de)-$approximation of the required shortest $od-$path.

On the other hand, for a given recursion budget $r\in \naturals$, it holds that $\s = \s(r) = \eps + \frac{\eps}{(1+\eps/\psi)^{r+1} - 1} = \frac{\eps\cdot(1+\eps/\psi)^{r+1}}{(1+\eps/\psi)^{r+1} - 1}$ is guaranteed by $\alg{RQA}$.
{\flushright\qed}
\end{proof}

Note that for time-independent, undirected-graphs (for which $\La_{\min} = \La_{\max}=0$ and $\zeta = 1$) it holds that $\psi = 2+\eps$. If we equip our oracle with \emph{exact} rather than $(1+\eps)-$approximate landmark-to-vertex distances (i.e., $\eps = 0$), then in order to achieve $\s = \de  = \frac{2}{t+1}$ for some positive integer $t$, our recursion budget $r$ is \emph{upper bounded} by $\frac{\psi}{\de}-1 = t$. This is exactly the amount of recursion required by the approach in \cite{Agarwal-Godfrey-2013} to achieve the same approximation guarantee.
That is, at its one extreme ($\La_{\min} = \La_{\max} = 0,~ \zeta = 1$, $\psi = 2$) our approach matches the bounds in \cite{Agarwal-Godfrey-2013} for the same class of graphs, without the need to grow balls from both the origin and destination vertices. Moreover, our approach allows for a \emph{smooth} transition from static and undirected-graphs to directed-graphs with FIFO arc-delay functions. The required recursion budget now depends not only on the targeted approximation guarantee, but also on the degree of asymmetry (the value of $\zeta\geq 1$) and the steepness of the shortest-travel-time functions (the value of $\La_{\max}$) for the time-dependent case. It is noted that we have recently become aware of an improved bidirectional approximate distance oracle for static undirected graphs \cite{Agarwal-2013} which outperforms \cite{Agarwal-Godfrey-2013} in the stretch-time-space tradeoff.

\subsection{Complexity of $\alg{RQA}$}
\label{subsection:RQA:complexity}

It only remains to determine the query-time complexity $\mathcal{Q}_{RQA}$ of $\alg{RQA}$. This is provided by the following theorem.
\begin{theorem}
\label{thm:RQA-query-time-complexity}
For the query-time complexity $\mathcal{Q}_{RQA}$ of $\alg{RQA}$ the following hold:
\begin{eqnarray*}
&& \Exp{\mathcal{Q}_{RQA}} \in O((1/\rho)^{r+1}\cdot\ln(1/\rho)\cdot\log\log(K_{\max}))\,.
\\
&& \Prob{\mathcal{Q}_{RQA}
	\in \Order{\left(\frac{\ln(n)}{\rho}\right)^{r+1} \left[\ln\ln(n)+\ln\left(\frac{1}{\rho}\right)\right] \log\log(K_{\max})}}
	\in 1 - \Order{\frac{1}{n}}\,.
\end{eqnarray*}
\end{theorem}
\begin{proof}
Recall that for any vertex $w\in V$ and any departure-time $t_w\in [0,T)$, the size of the outgoing $\alg{TDD}-$ball $B_w = B[w](t_w)$ centered at $(w,t_w)$ until the first landmark vertex is settled, behaves as a geometric random variable with success probability $\rho \in (0,1)$. Thus, $\Exp{|B_w|} = \frac{1}{\rho}$ and $\forall \beta\in \naturals,~ \Prob{|B_w| > \beta} \leq \exp(-\rho\cdot \beta)$. By applying the trivial union bound, one can then deduce that:
\(
	\forall W\subseteq V, \Prob{\exists w\in W: |B_w| > \beta}
	\leq |W|\exp(-\rho\beta)  = \exp\left(-\rho\beta + \ln(|W|)\right)\,.
\)

Assume now that we somehow could guess an upper bound $\beta^*$ on the number of vertices in every ball grown by an execution of $\alg{RQA}$. Then, since the out-degree is upper bounded by $2$, we know that the boundary $\partial B$ of each ball $B$ will have size $|\partial B|\leq 2 |B|$. This in turn implies that the branching tree that is grown in order to implement the ``guessing'' of step 7.1 in $\alg{RQA}$ (cf.~Figure~\ref{fig:RQA-pseudocode}) via exhaustive search, would be bounded by a complete $(2\beta^*)-$ary tree of depth $r$. For each node in this branching tree we have to grow a new $\alg{TDD}-$ball outgoing from the corresponding center, until a landmark vertex is settled. The size of this ball will once more be upper-bounded by $\beta^*$.
Due to the fact that the out-degree is bounded by $2$, at most $2\beta^*$ arcs will be relaxed. Therefore, the running time of growing each ball is $\Order{\beta^*\ln(\beta^*)}$. At the end of each $\alg{TDD}$ execution, we query the oracle for the distance of the newly discovered landmark to the destination node. This will have a cost of $\Order{\log\log((K^*+1)\cdot U)}$, where $U$ is the maximum number of required breakpoints between two concavity-spoiling arc-delay breakpoints in the network, since all the breakpoints of the corresponding shortest-travel-time function are assumed to be organized in a predecessor-search data structure.  The overall query-time complexity of $\alg{RQA}$ would thus be bounded as follows:
\begin{eqnarray*}
	\mathcal{Q}_{RQA}
	&\leq & \frac{(2\beta^*)^{r+1} - 1}{2\beta^*-1}\cdot\Order{\beta^*\ln(\beta^*)+\log\log((K^*+1)\cdot U)}
	\\
	&\in & \Order{(\beta^*)^{r+1}\ln(\beta^*) + \beta^{r}\log\log((K^*+1)\cdot U)}
\end{eqnarray*}
Assuming that $\log\log((K^*+1)\cdot U) \in \Order{\beta^*\log(\beta^*)}$, we have that 	
\(
	\mathcal{Q}_{RQA} \in \Order{(\beta^*)^{r+1}\ln(\beta^*)}\,.
\)
If we are only interested on the expected running time of the algorithm, then each ball has expected size $\Order{\frac{1}{\rho}}$ and thus $\Exp{\mathcal{Q}_{RQA}} \in \Order{\left(\frac{1}{\rho}\right)^{r+1} \ln\left(\frac{1}{\rho}\right)}$.

In general, if we set $\beta^* = \frac{r\ln(n)}{\rho}$, then we know that $\alg{RQA}$ will grow
$|W|\in\Order{\left(\frac{r\ln(n)}{\rho}\right)^{r}}$ balls, and therefore:
\begin{eqnarray*}
	\Prob{\forall w\in W, |B_w|\leq \frac{r\ln(n)}{\rho}}
	&\geq& 1 - \exp\left( -\rho \frac{r\ln(n)}{\rho} + r\cdot[\ln\ln(n) +\ln(1/\rho)]\right)
	\\
	&\in& 1 - \Order{\frac{1}{n}}
\end{eqnarray*}
Thus, we conclude that:
\[
	\Prob{\mathcal{Q}_{RQA}
	\in \Order{\left(\frac{\ln(n)}{\rho}\right)^{r+1}\cdot\left[\ln\ln(n)+\ln\left(\frac{1}{\rho}\right)\right]}}
	\in 1 - \Order{\frac{1}{n}}\,.
\]
{\flushright\qed}
\end{proof}

\section{Main Results}
\label{section:main-results}

In this section, we summarize the main result of our paper and establish the tradeoff between preprocessing, query time and stretch. Recall that $U$ is the worst-case number of breakpoints for an $(1+\eps)-$approximation of a \emph{concave} $(1+\eps)$-approximate distance function stored in our oracle, and $TDP$ is the maximum number of time-dependent shortest path probes during their construction\footnote{As it is proved in
Theorem~\ref{thm:SO-BISECTION-performance}, $U$ and $TDP$ are independent of the network size $n$.}. The following theorem summarizes our main result.
\begin{theorem}
\label{thm:flat-oracle-summary}
For sparse time-dependent network instances compliant with Assumptions \ref{assumption:Bounded-Travel-Time-Slopes} and \ref{assumption:Bounded-Opposite-Trips}, a distance oracle is provided with the following characteristics:
(a) it selects among all vertices, uniformly and independently with probability $\rho$, a set of landmarks;
(b) it stores $(1+\eps)-$approximate distance functions (summaries) from every landmark to all other vertices;
(c) it uses a query algorithm equipped with a recursion budget (depth) $r$.
Our time-dependent distance oracle achieves the following expected complexities:\\
(i) preprocessing space $\Order{\rho n^2 (1+K^*) U}$;\\
(ii) preprocessing time $\Order{\rho n^2 (1+K^*)\log(n) \log\log(K_{\max}) TDP}$;\\
(iii) query time $\Order{\left(\frac{1}{\rho}\right)^{r+1} \log\left(\frac{1}{\rho}\right) \log\log(K_{\max})}$.\\
The guaranteed stretch is $1 + \eps \frac{(1+\frac{\eps}{\psi})^{r+1}}{(1+\frac{\eps}{\psi})^{r+1}-1}$,
where $\psi$ is a fixed constant depending on the characteristics of the arc-travel-time functions, but is independent of the network size.
\end{theorem}
\begin{proof}
Immediate consequence of Theorems~\ref{thm:preprocessing-complexity}, \ref{thm:RQA-approximation-guarantee}, and \ref{thm:RQA-query-time-complexity}.
\qed
\end{proof}

Note that, apart from the choice of landmarks, our preprocessing and query algorithms are deterministic. The following theorem expresses explicitly the tradeoff between subquadratic preprocessing, sublinear query time and stretch of the proposed oracle.
\begin{theorem}
\label{thm:preprocessing-vs-query-tradeoff}
Let $G=\left(V,A,(D[a])_{a\in A}\right)$ be a sparse time-dependent network instance compliant with Assumptions \ref{assumption:Bounded-Travel-Time-Slopes} and \ref{assumption:Bounded-Opposite-Trips}.
Assume that our distance oracle is deployed on $G$ for:
(a) creating the landmark set uniformly at random with probability $\rho = n^{-a}$, for some $a\in\left(0,\frac{1}{r+1}\right)$;
(b) computing with the $\alg{BISECTION}$ method the $(1+\eps)$-approximate distance summaries for landmark-to-vertex distances;
(c) running $\alg{RQA}$ to respond to arbitrary queries $(o,d,t_o)\in V\times V\times[0,T)$, with approximation guarantee $1+\g\eps$, for some constant
\(
		\g = \frac{(1 + \eps/\psi)^{r+1}}{(1 + \eps/\psi)^{r+1} - 1} > 1\,.
\)
Provided that the degree of disconcavity of $G$ is $K^* \in \polylog(n)$, then the following expected complexities hold:
\begin{eqnarray*}
\mbox{Preprocessing space:}\TAB	& \Exp{S}	& \in \tildeOrder{n^{2-a}}
\\
\mbox{Preprocessing time:}\TAB 	& \Exp{P}	& \in \tildeOrder{n^{2-a}}
\\
\mbox{Query time:}\TAB 			& \Exp{Q}	& \in \tildeOrder{n^{(1+r)a}}
\end{eqnarray*}
\end{theorem}
\begin{proof}
Since we have assumed that $K^*\in \polylog(n)$ and since by Theorem~\ref{thm:SO-BISECTION-performance}
$U$ and $TDP$ are independent of the network size $n$ and hence can be treated as constants,
it follows from Theorem~\ref{thm:flat-oracle-summary} that
the expected preprocessing space and time complexities, $\Exp{S} \in \tildeOrder{n^{2-a}}$ and $\Exp{P} \in \tildeOrder{n^{2-a}}$,
are indeed subquadratic. It similarly follows from the same theorem that the expected query time
is $\Exp{Q_{\alg{RQA}}} \in \tildeOrder{n^{(1+r)a}}$. Hence, to complete the proof is remains to show
that $\tildeOrder{n^{(1+r)a}}$ is also sublinear, i.e., $(r+1)a < 1$. Recall that, by Theorems~\ref{thm:flat-oracle-summary} and \ref{thm:RQA-approximation-guarantee}, a $(1+\g\eps)$-approximate solution is returned, for
\(
	\g = \frac{(1 + \eps/\psi)^{r+1}}{(1 + \eps/\psi)^{r+1} - 1}
\)
which holds by our assumption. Alternatively, to assure a desired approximation guarantee $1+\g\eps$ for arbitrary queries, value $\g>1$ and a given approximation guarantee $1+\eps$ for the preprocessed distance summaries, we should set appropriately the recursion budget to
\[
	r = \frac{\log\left(\frac{\g}{\g-1}\right)}{\log\left(1+\frac{\eps}{\psi}\right)} - 1
\]
\qed
\end{proof}
%

\section{Approximate Shortest Path Reconstruction}
\label{sec:path-reconstruction}

As it is customary in the distance oracles literature, the query-time complexities of our algorithms concern only the determination, for a given query $(o,d,t_o)\in V\times V\times[0,T)$, of an upper bound $\De[o,d](t_o)$ on the shortest-travel-time $D[o,d](t_o)$, or equivalently an arrival-time $\tau_d := t_o+\De[o,d](t_o)$ at $d$ which guarantees this upper bound on the travel-time.

Our goal in this section is to describe a method for reconstructing an actual $od-$path (roughly) guaranteeing this travel-time bound, in time (additional to the already reported query-time) that is roughly linear in the number of its constituent arcs. Indeed, our goal is only to exploit the precomputed landmarks-to-vertices approximate distance summaries, along with the value $\tau_d$ that was computed on the fly, in order to discover such a path. Indeed, the origin-to-landmark path is computed ``on-the-fly'' and the main challenge is to construct the remaining landmark-to-destination approximate path that would guarantee the reported arrival-time at the destination. A natural approach would be to mimick the path reconstruction from the destination back to the landmark, based only on the (upper bound on the) arrival-time at the destination, as is typically done in the time-independent case. This would indeed be possible, if we had at our disposal \emph{exact} landmark-to-vertices distance summaries. But we can only afford for $(1+\eps)-$approximate distance summaries of the actual travel-time functions and thus the only thing we know is that $\tau_d\in t_o + D[o,d](t_o) \cdot[1 , 1+\eps]$. Thus, we cannot be sure that such a reconstruction is indeed possible: It might be the case that $\tau_d = t_d := t_o + D[o,d](t_o)$ while at the same time some of the approximate distances from the landmark to intermediate vertices along the path are indeed inexact.

To resolve this issue, we shall exploit the fact that the approximate distance summaries created during preprocessing, correspond to travel-time functions along a shortest-paths tree from the landmark to all possible destinations, for the given departure time. This tree is actually a valid \emph{approximate} shortest paths tree, not only for the sampled departure time, but also for the entire time-interval of departures till the next sample point. Due to the sparsity of the graph, we can be sure that only a constant number of bits is required per breakpoint in the pwl-approximations, in order for each vertex to memoize its own parent in such a tree (as a function of the departure-time from the landmark). The path reconstruction is then conducted by moving from the destination towards the landmark, evaluating the right leg of the corresponding approximate distance summary in each intermediate vertex, so that the appropriate parent (and the latest departure time from it) is selected. In overall, the construction time will be almost linear in the number of arcs constituting the required approximate shortest path (times an $\Order{\log\log(K_{\max})}$ factor, for evaluating the right leg in the approximate distance summary of each intermediate vertex).

\section{Conclusions}
\label{section:conlusions}

We have presented the first time-dependent distance oracle for sparse networks, compliant with Assumptions \ref{assumption:Bounded-Travel-Time-Slopes} and \ref{assumption:Bounded-Opposite-Trips}, that achieves subquadratic preprocessing space and time, sublinear query time, and stretch factor arbitrarily close to 1.
Our approach is based on a new algorithm, built upon the bisection method, that computes one-to-all approximate distance summaries from a set of selected landmarks to all other vertices of the network as well as on a new recursive query algorithm. Our assumptions, justified by an experimental analysis of real-world and benchmark data, allow us to achieve a smooth transition, from the undirected (symmetric) and static world to the directed (asymmetric) and time-dependent world, through two parameters that quantify the degree of asymmetry ($\zeta$) and its evolution over time (expressing the steepness of the shortest travel-time functions via $\La_{\min}$ and $\La_{\max}$).

It would be quite interesting to come up with a new method for computing approximate distance summaries, that avoids the dependence of the preprocessing complexities on the number $K^*$ of concavity-spoiling breakpoints.

Finally, almost all distance oracles with provable approximation guarantees in the literature, even for the static case, 
target at sublinearity in query times with respect to the network size. 
A very important aspect would be to propose query algorithms that are indeed sublinear not only in worst-case, 
but also sublinear on the Dijkstra rank of the destination vertex. 

\bibliographystyle{plain}


\end{document}